\documentclass[preprint,12pt]{elsarticle}

\usepackage{graphicx}
\usepackage{amssymb}

\usepackage{lineno}

\usepackage{float}
\usepackage{amsmath}
\usepackage{nicefrac}
\usepackage{subcaption}
\usepackage{textcomp}

\begin{document}

\begin{frontmatter}


\title{Compositional Pore-Network Modeling of Gas-Condensate Flow}

\author{P. K. P. Reis}
\ead{paulareis@lmmp.mec.puc-rio.br}

\author{M. S. Carvalho \corref{cor}}
\ead{msc@puc-rio.br}

\cortext[cor]{Corresponding author}



\address{Department of Mechanical Engineering, Pontif\'{\i}cia Universidade Cat\'olica do Rio de Janeiro, Rio de Janeiro-RJ 22451-900, Brazil}

\begin{abstract}
Liquid dropout and retention in gas-condensate reservoirs, specially in the near wellbore region, obstruct gas flowing paths and impact negatively the produced fluid volume and composition. Yet, condensate banking forecasting is commonly inaccurate, as experiments seldom reproduce reservoir extreme conditions and complex fluid composition, while most pore-scale models oversimplify the physics of phase transitions between gas and condensate. To address this gap, a fully implicit isothermal compositional pore-network model for gas and condensate flow is presented. The proposed pore-networks consist of 3D structures of pores connected by constricted circular capillaries. Hydraulic conductances are calculated for the capillaries, which can exhibit single-phase flow or two-phase annular flow, according to local gas and liquid saturations, or be blocked by a liquid bridge, when capillary forces overcome viscous forces. A PT-flash based on the Peng-Robinson EoS is performed at control volumes defined for the pores at each time step, updating the phases properties. Flow analyses were carried based on coreflooding experiments reported in the literature, with matching fluid composition and flow conditions, and approximated pore-space geometry. Predicted and measured relative permeability curves showed good quantitative agreement, for two values of interfacial tension and three values of gas flow velocity.

\end{abstract}

\begin{keyword}
Gas Condensate Relative Permeabilities \sep  Dynamic pore-network Modeling \sep Compositional Modeling

\end{keyword}

\end{frontmatter}

\section{Introduction}
\label{S:1}
Hydrocarbon recovery from gas condensate reservoirs tends to undergo a sharp decline in the event of condensate banking \citep{Banbi2000,barnum1995gas}. As the near wellbore region is depleted below the gas dew point pressure, retrograde condensation occurs, leading to a subsurface liquid saturation buildup. Consequently, not only the gas flow rate is significantly hindered, but valuable heavier components are trapped in the porous medium, leading to a leaner produced gas.

In order to prevent low productivity and promote economic viability in gas condensate field developments, several production optimization and EOR methods have been proposed recently \citep{azin2019production,davarpanah2019simulation,hoseinpour2019condensate,mazarei2019feasibility,zhang2020investigation}. These methods commonly involve the injection of gas ($CO_{2}$, $N_{2}$, $CH_{4}$, $C_{2}H_{6}$ and re-injected
gas) to alter phase behavior and sustain reservoir pressure, and the injection of wettability modifier chemicals to alter rock surfaces from condensate-and water-wet to preferentially gas-wet. Adequate development of such production enhancement schemes requires deep understanding of the coupled flow of gas and condensate in porous media and its dependency on parameters like flow rate, phases properties, pore morphology and wettability, which has been proved to be fundamentally different from that observed in conventional two-phase flow systems \citep{jamiolahmady2003}.

To this end, coreflooding experiments have been widely used and provided relevant macro-scale data, such as critical condensate saturation ($S_{cc}$) and relative permeability curves, as well as their dependency on interfacial tension and flow rate. \citep{Chen1995,kr_rate_2,kr_rate_1,mott2000}. Additionally, visualization of gas and condensate flow in micromodels contributed with valuable qualitative data regarding flow patterns, phase change and phase trapping in pores \citep{Harassi2009,Coskuner1997,Dawe2007}. However, due to experimental operation constraints, these experiments seldom reproduce extreme pressure and temperature conditions or use complex fluid compositions normally occurring in gas condensate reservoirs. This leads to significant uncertainties in results, since model fluids do not always accurately capture the flow characteristics of compositionally complex reservoir fluids and may not mimic the reservoir wettability \citep{nagarajan2004comparison}.

As an alternative, similar data can be obtained via pore-scale modeling \citep{blunt2013pore}. Pore-network models of multiphase flow could be particularly suitable for this purpose, as they have shown promising results for predictive purposes, while being computationally less demanding than direct models \citep{joekar2012analysis}. In the past decades extensive research effort has been directed to the development of this class of simulation tools, with a few regarding specifically gas and condensate flow. An overview of the works available in the literature concerning pore-network modeling of gas and condensate flow is presented next.

\citet{mohammadi1990} represented the pore space as Bethe trees and calculated relative permeabilities using percolation theory. Threshold radii as a function of pressure were defined to determine which capillaries were filled with liquid. Also, condensate accumulation in tight corners was considered by adopting capillaries with convex polyhedral cross sections. Relative permeabilities showed sensitivity to pore geometry, connectivity and size distribution.  \citet{fang1996} developed a phenomenological model to calculate condensate critical saturation and its dependency on interfacial tension and contact angle hysteresis. Their model contemplated only gravitational and capillary forces and the network was composed of cylindrical capillaries. At each time step, a finite amount of condensate was arbitrarily added to the capillaries with radii below the threshold radius of $20\mu m$. Critical condensate saturation increased as interfacial tension rose and it was also affected by contact angle hysteresis. \citet{wang1999} also developed a model for $S_{cc}$ determination. They used a 3D pore-network with converging-diverging pore throats, with square cross section, so that liquid accumulation in the corners was assessed. Condensation in the pore throats was also associated with their radii. Under gravitational and capillary forces, they have observed that $S_{cc}$ increased with interfacial tension and decreased with connate water saturation. The same authors upgraded their model by including viscous forces and relative permeability calculations \citep{wang2000}. In this work, two scenarios were analyzed: low capillary number regime and low condensate saturation/high pressure gradient flow regime. In the first, relative permeabilities were strongly sensitive to the pore structure. In the second, both gas and condensate relative permeabilities increased with their saturation and pressure gradient. \citet{li2000} also included viscous forces in a model based on \cite{fang1996} and calculated both $S_{cc}$ and relative permeabilities. They have concluded that the effect of gravity and viscous forces in the $S_{cc}$ is significant especially at low interfacial tension, but that the most relevant parameter is wettability. As for the relative permeabilities, interfacial tension and viscous forces are important parameters, which is in line with experimental data. \citet{jamiolahmady2000} developed a mechanistic model for gas and condensate flow in a single pore throat to describe the rate effect on relative permeabilities. In this model, both phases are injected in a conical pore and the fluid spacial distribution is controlled by an evolution equation that describes the condensate film position, used when the flow is annular, and a gas advancement equation, used when a bridge of condensate is formed. The positive effect of flow rate on the gas relative permeability was more pronounced than that on the condensate relative permeability, while a decrease in interfacial tension led to a increase in the permeabilities of both phases. The same authors also developed a 3D pore-network model and compared results with coreflooding experiments \citep{jamiolahmady2003}. The network was composed of cylindrical capillaries and the evolution equation \citep{jamiolahmady2000} was replaced by a simpler evolution time correlation. The predicted values were not quantitatively equivalent to the corresponding measured ones, but the rate sensitivity of the relative permeabilities was similar to the measured data. \citet{bustos2003} developed a mechanistic model with 2D and 3D networks of capillaries with square cross section. Similarly to \cite{fang1996}, condensation happened in discrete steps, but in this case it started at the capillary corners before evolving into a condensate bridge. Viscous forces were not included and relative permeability, notably for the gas, was sensitive to geometric parameters.

Those works have in common the shortcoming of not reproducing the effects of phase equilibrium thermodynamics on gas and condensate coupled flow. Important phenomena like the accumulation of heavier hydrocarbons in the porous media or the dependency of interfacial tension and saturation on pressure are not appraised. In order to address this issue, \citet{momeni20173d}, have proposed a compositional 3D pore-network model for dynamic displacement of gas and condensate in wellbore region. In their model, a flash calculation is performed at each throat at every time step, determining the phases properties and saturations. Square capillaries were used and two fluid configurations were allowed: gas core with condensate flowing in the corners or a fully saturated throat by condensate. The pores were considered to be volumeless and impose no restriction to flow, while the throats conductances were calculated with Poiseuille law with no slip boundary condition at the solid-fluid interfaces and free slip at the fluid-fluid interfaces. Fluids were injected at the inlet at fixed flow rate and separated mass balances for each phase at each pore were used to calculate the pressure fields. Relative permeabilities were calculated during unsteady state flow and were compared to experimental coreflooding data. Although compositional, this model has some considerable limitations. The pressure determination in the pores is based solely on mass balances, not coupled with volume consistency equations and molar balance for each component. Also, free slip in the fluid-fluid interface is more suited for two-phase systems where the fluids have significantly different viscosities, and, finally, relative permeabilities were calculated during unsteady flow, where the outlet flow rates could be significantly different than inlet imposed flow rates. These drawbacks were addressed by \citet{santos}. They have represented porous media as 2D regular grids of straight cylindrical capillaries, where the condensate and gas flow in annular configuration. The phases velocity profiles were obtained with a steady solution of Navier-Stokes equation. Molar balance equations coupled with volume consistency equations were solved at each time step, providing molar content and pressure for every pore. Local thermodynamic equilibrium was enforced in each pore and phase equilibrium calculations used the \citet{Peng:1976} equation of state. Their work was limited to high capillary number regime and, consequently, did not include the formation of bridges of condensate in the capillaries nor capillary forces, which prevented the evaluation of the relative permeabilities dependency on flow rate and interfacial tension.

In this work, capillary pressure, more realistic network geometry and new pore-level phase distribution comprising the snap-off phenomenon are included in the compositional pore-network model presented by \citet{santos}. With the proposed model, relative permeability curves for gas and condensate were generated for three different flow velocities and two interfacial tension values, and results showed good quantitative agreement with experimental data. 

\section{Pore-network  Model}
\label{CPNM}
\subsection{Pore-network Geometry}

The pore-networks used in the present analysis consist of 3D structures of pores connected by circular constricted capillaries. Pores are identified by the index \textit{i} and capillaries by the index \textit{j}. The connectivity between them is mapped by an incidence matrix, \textbf{C}, with elements defined as follows: 

\begin{equation*}
\label{im}
c_{ij} =
  \begin{cases}
     \;\;\,0 & \text{, if the capillary \textit{j} is not connected to the pore \textit{i}}  \\
    -1 &  \text{, if the capillary \textit{j} enters the pore \textit{i}} \\
    +1 &  \text{, if the capillary \textit{j} exits the pore \textit{i}} 
  \end{cases}
\end{equation*}

\begin{figure}[H]
    \centering
    \includegraphics[width=60mm]{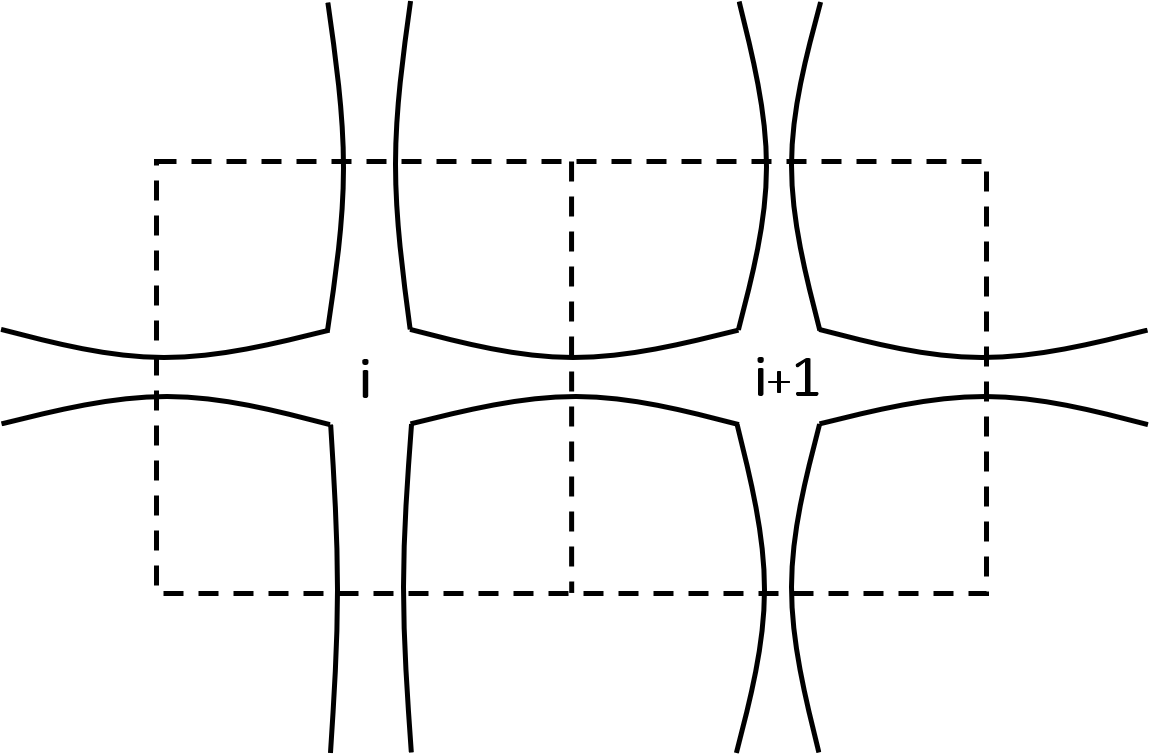}
    \caption{Definition of pore volume}
    \label{fig:volumeporo}
\end{figure}

 Capillaries are characterized by a length $L_j$ and a hyperbolic profile (eq. \ref{r_fun}), with two geometric parameters, the maximum radius, $R_{max,j}$ and the throat radius, $R_{min,j}$ . Pores are defined by a control volume enclosing half the length of the capillaries contiguous to them, as illustrated in Figure \ref{fig:volumeporo}.

 \begin{subequations}
\label{r_fun}
\begin{align}
 r_j(x) = \sqrt{a_j+b_jx^2}  \\
 a_j =  R_{min,j}^2    \\
 b_j =  \left(\frac{2}{L_j}\right)^2(R_{max,j}^2-R_{min,j}^2)
 \end{align}
\end{subequations}

The spacial configuration of pores and capillaries can be defined based on 3D pore-scale imaging of real rock samples, which leads to networks with variable capillary lengths and coordination number (number of capillaries connected to each pore). Alternatively, the network can be constructed based on 3D regular lattices, with realistic pore and throat size distributions, in the absence of more detailed petrophysical data.

\subsection{Two-Phase Flow in the Capillaries}
\label{TPF}
\subsubsection{Flow Patterns Overview}
\begin{figure}[H]
    \centering
    \includegraphics[width=50mm]{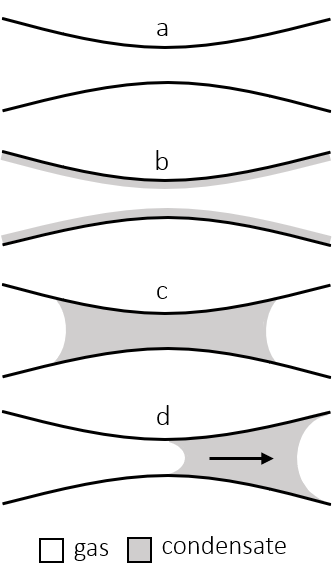}
    \caption{Evolution of condensate configuration in a throat}
    \label{fig:cond_ev}
\end{figure}

For a gas condensate reservoir, the flow at pressures above the dew point pressure ($P_{dp}$) contains only the gas phase (Fig.\ref{fig:cond_ev}a). If the pressure level is lowered bellow the $P_{dp}$, however, a liquid phase emerges. It is considered in the model that the condensate wets completely the capillary walls (contact angle $\theta\approx 0$) and tends to flow adjacent to them, developing an annular flow pattern (Fig.\ref{fig:cond_ev}b).

The annular flow in a capillary may be interrupted nonetheless, if the condensate film  thickness ($t$) reaches a critical value ($t_{crit}$) above which the annular configuration becomes unstable. In this case, the liquid tends to evolve into a bridge accommodated at the capillary mid-section, in a phenomenon known as snap-off (Fig.\ref{fig:cond_ev}c). While a capillary accommodates a condensate bridge, the flow of gas and condensate is obstructed. After the snap-off, the flow can be reestablished if the pressure drop between the capillary extremities exceeds a critical capillary pressure value, and the liquid bridge is moved (Fig.\ref{fig:cond_ev}d).

\subsubsection{Gas and Condensate Annular Flow Conductances}

The volumetric flow rate of gas and condensate in a capillary can be determined using the concept of fluid conductance: $g=\nicefrac{q}{\Delta P}$, where $\Delta P$ is the pressure difference in the capillary. The conductance calculations in this model are based on the steady-state solution of the Navier-Stokes equation for a laminar two-phase annular flow in a straight capillary, given by equations \ref{G} \citep{santos}.

\begin{subequations}
\label{G}
\begin{align}
 g_{g} &=\frac{\pi}{8\mu _{g} L}(S_{g} R^2)^2+\frac{\pi}{4\mu _{l} L}(R^2-S_{g} R^2) S_{g} R^2  \\
 g_{l} &=\frac{\pi}{8\mu _{l} L}(R^2-S_{g} R^2)^2 
 \end{align}
\end{subequations}

In equations \ref{G}, $\mu _g$ and $\mu _{l}$ are the viscosities of the gas and liquid phases, $S_{g}$ is the gas saturation and $R$ and $L$ are, respectively, the equivalent radius and length of the capillary. In order to adapt these conductances for constricted capillaries, an equivalent constant radius $R$ was calculated for each capillary in the network. $R$ represents the radius of a straight capillary that would yield the same volumetric flow rate as the constricted capillary, when subjected to the same pressure difference. The volumetric flow rate as a function of the pressure drop for a laminar, steady-state flow of a Newtonian fluid in an axisymmetric tube with hyperbolic profile \cite{req} was used, leading to the Equation \ref{r_eq} for the equivalent radius.

\begin{equation}
\label{r_eq}
R=\left[ \frac{2R_{min}^3R_{max}^2\sqrt{R_{max}^2-R_{min}^2}}{R_{min}\sqrt{R_{max}^2-R_{min}^2}+R_{max}^2\arctan\left( \sqrt{\frac{R_{max}^2-R_{min}^2}{R_{min}^2}} \right)} \right]^{\nicefrac{1}{4}}
\end{equation}

\subsubsection{Capillary Blocking Conditions}
\label{cap_block}
Several criteria for the formation of the condensate bridge in capillaries, in the context of pore-network modeling, have been suggested in the literature. These criteria are commonly highly simplified, so that the model computational effort is reduced. A condition based simply on a capillary radius threshold of $20\mu m$ was adopted by \citet{fang1996}. \citet{bustos2003} proposed a model containing square capillaries in which the condensate accumulation was initially contained in the capillary corners and the bridge of liquid was formed at condensate saturations greater than 21.4\%. This value represents the condensate saturation at which the contact between the gas phase and the pore walls is lost. The same value was adopted by \citet{momeni20173d}. \citet{jamiolahmady2003} suggested a criterion based on the time required for a bridge to be formed, as a function of the capillary geometry, fluid properties and initial condensate film thickness. In our model, it was considered that a condensate bridge is formed upon reaching a critical condensate film thickness, proportional to the constriction radius of each capillary. Additionally, for the snap-off to happen, the geometry-controlled condition for the fluid break-up in constricted capillaries proposed by \citet{beresnev2009condition} had to be met. This criterion, shown in Equation \ref{berev}, reduces to the condition for the occurrence of the Plateau-Rayleigh instability in the limiting case of cylindrical capillaries. 

\begin{equation}
\label{berev}
L>2\pi\sqrt{(R_{min}-t)(R_{max}-t)} 
\end{equation}

For $t_{crit}=0.25R_{min}$, (which will be adopted in the analysis presented in this paper) the condensate saturations at which the snap-off occurs vary from $\approx$ 1\%, for capillaries with very high $R_{max}/R_{min}$ ratio, to 43.75\%, for unconstricted capillaries. This interval is consistent with the values presented in the literature.

\begin{figure}[H]
    \centering
    \includegraphics[width=65mm]{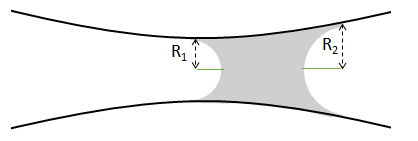}
    \caption{Condensate bridge radii, $R_1$ and $R_2$, used to calculate $\Delta P_{crit}$}
    \label{fig:dp_crit}
\end{figure}

Once blocked, the critical pressure drop value for a capillary to be unobstructed is given by Equation \ref{dP_crit}. It is a function of the interfacial tension, $\sigma$, between the phases and the radii of the condensate bridge meniscii, $R_1$ and $R_2$. The interfacial tension is calculated with the correlation proposed by \citet{WK}, and the radii are related to the capillary geometry and the condensate saturation, as depicted in Figure \ref{fig:dp_crit}.

\begin{equation}
\label{dP_crit}
\Delta P_{crit}=2\sigma \left(\frac{1}{R_{1}}-\frac{1}{R_{2}}\right)
\end{equation}

The switch from opened to closed states in a capillary was implemented in the model by multiplying the conductances (eq. \ref{G}) by a continuous approximation of a unit step function $H$, presented in \ref{app:H}. $H$ is equal to unity, when the conditions for annular flow are met, and equal to zero, when the flow is blocked by a condensate bridge.

\subsection{Equations Governing the Flow}
\label{NLS}

Once the capillary conductances have been defined, the fluid content and pressure in the pores can be calculated. This is done via the coupled solution of Molar Balance Equations (section \ref{MBE}) and Volume Consistency Equations (section \ref{VCE}), along with the appropriate boundary conditions (section \ref{BC}). The resulting system of non linear equations, described in section \ref{NRM}, relates the model variables, $P_i$, $N_i^k$ and $s_i$, respectively, the pressure and the number of moles of each component $k$, for all pores, and a source/sink term, for the pores located at the inlet/outlet of the network, and is solved using the Newton Raphson Method.

\subsubsection{Molar Balance Equation}
\label{MBE}
The  fluid content in a pore depends on the molar flow rates of the capillaries connected to it. Therefore, the molar balance equation (eq. \ref{mbeq}) of each component $k$ in a pore $i$ is written in terms of the molar flow rate, $\dot{n}_j^k$, through its adjacent capillaries. Also, the term $s_i^k$ is used for the pores located at the inlet and outlet of the network, to account for the molar flow through its boundaries. The molar flow in a capillary, given by Equation \ref{nj}, converts the volumetric flow, calculated with the conductances and the pressure drop, into the molar flow, using the molar fraction of each component in the gas and liquid phases, $y^k$ and $x^k$, and their molar densities, $\xi_g$ ans $\xi_l$. Equation \ref{dpj} represents the pressure that drives the flow through the capillaries. It amounts to the difference in the pressures of the pores connected by the capillary, minus the interfacial pressure difference, when a condensate slug is formed. The inclusion of the interfacial pressure difference in eq. \ref{dpj} is controlled by the function $H^{int}$, discussed in \ref{app:H}.

\begin{equation}
\label{mbeq}
\frac{\partial N^k_i}{\partial t}=-\displaystyle\sum_{j=1}^{n_{edge}} c_{ij}\dot{n}_j^k+s_i^k
\end{equation}

\begin{equation}
    \label{nj}
    \dot{n}_j^k=H_j(y^k\xi _{g}g_{g}+x^k\xi _{l}g_{l})_j\Delta P_j
\end{equation}

\begin{equation}
    \label{dpj}
    \Delta P_j=\displaystyle\sum_{m=1}^{n_{node}} c_{mj}P_m - H_j^{int}\Delta P_j^{int}
\end{equation}

\subsubsection{Volume Consistency Equation}
\label{VCE}
In the proposed model, the pores contain only the gas and condensate phases. The volume of each pore, therefore, can be written as: $V_i=V_i^{g}+V_i^{l}$. Given that the network is slightly compressible, the pore volume can also be written as a function of the pore pressure and compressibility $\nu _i=\frac{1}{Vi}\frac{\partial V_i}{\partial P_i}$. 

Considering $\overline{V_i}$ the volume of a pore at a reference pressure $\overline{P_i}$, its volume at any pressure can be approximated by Equation \ref{vceqs}a. In regard to the pore contents, $V_i^{g}+V_i^{l}$ can be determined by relating $P_i$ and $T$ with the fluid parameters $\mathcal{L}_i$, $Z_i^{g}$, $Z_i^{l}$, $x_i^k$ and $y_i^k$ (respectively: the fraction of the $N_i$ moles in the liquid phase,  the compressibility factors and the molar fractions of each component $k$ in the gas and liquid phases), as presented in Equation \ref{vceqs}b.

\begin{subequations}
\label{vceqs}
\begin{align}
V_i &\approx \overline{V_i}[1+\overline{\nu _i}(P_i-\overline{P})]
\\
V_i &=N_i\left[\mathcal{L}_i\left(\frac{Z_i^{l}RT}{P_i}-\displaystyle\sum_{k=1}^{n_c} v_kx_i^k\right)+(1-\mathcal{L}_i)\left(\frac{Z_i^{g}RT}{P_i}-\displaystyle\sum_{k=1}^{n_c} v_ky_i^k\right)\right]
 \end{align}
\end{subequations}

The combination of the equations above leads to the volume consistency equation, given by Equation \ref{vceq}.

\begin{equation}
    \label{vceq}
    N_i-\frac{\overline{V_i}[1+\overline{\nu _i}(P_i-\overline{P})]}{\mathcal{L}_i\left(\frac{Z_i^{l}RT}{P_i}-\sum_{k=1}^{n_c} v_kx_i^k\right)+(1-\mathcal{L}_i)\left(\frac{Z_i^{g}RT}{P_i}-\sum_{k=1}^{n_c} v_ky_i^k\right)}=0
\end{equation}

\subsubsection{Boundary Conditions}
\label{BC}

Two sets of boundary conditions can be chosen for a flow analysis in the model. In the first, pressure is fixed at both inlet and outlet faces of the network, while in the second, molar flow rate is fixed at the inlet and pressure at the outlet. Other parameters that have to be imposed are the composition of the fluid injected in the network and the temperature.

\subsubsection{Solution via Newton Raphson Method}
\label{NRM}
Equations \ref{mbeq} and \ref{vceq}, associated with the appropriated boundary conditions, form the system of non-linear equations that relate the variables $N_i^k$, $P_i$ and $s_i$ at each time step. The solution is obtained using Newton-Raphson method, where the unknown vector is \textbf{u=}$[N_i^k; P_t; s_i]$ and the residue vector is \textbf{R=}$[(R_N)_i^k; (R_V)_i; (R_C)_i]$. The entries of the vector \textbf{R} are presented in equations \ref{NR}. 

\begin{subequations}
\label{NR}
\begin{align}
\begin{split}
 &(R_N)_i^k =(N_i^k)^{t+1}-(N_i^k)^{t}+ \\ 
 & dt\left\{ \displaystyle\sum_{j=1}^{n_{edge}} c_{ij}[H_j^\tau(y^k\xi_{g} g_{g}+x^k\xi_{l} g_{l})_j^\tau \left( \displaystyle\sum_{m=1}^{n_{node}} c_{mj}P_m^{t+1} - H_j^{int,\tau}\Delta P_j^{int,\tau}\right) -z_{si}^{k\tau} s_i^{t+1}]\right\}  
 \end{split} \\
 &(R_V)_i =(N_i)^{t+1}-\frac{\overline{V_i}[1+\overline{\nu_i}(P_i^{t+1}-\overline{P})]}{\mathcal{L}_i^\tau\left(\frac{(Z_i^{l})^\tau RT}{P_i^{t+1}}-\sum_{k=1}^{n_{c}} v_k(x_i^k)^\tau\right)+(1-\mathcal{L}_i^\tau)\left(\frac{(Z_i^{g})^\tau RT}{P_i^{t+1}}-\sum_{k=1}^{n_{c}} v_k(y_i^k)^\tau\right)} \\
 &(R_C)_i =
\begin{cases}
dt\frac{(s_i-s_i^{imp})}{RT}\text{ ,} & \text{for imposed molar flow at node \textit{i}}\\
\frac{(P_i-P_i^{imp})\overline{V_i}}{RT}\text{ ,}   & \text{for imposed pressure at node \textit{i}}
\end{cases}
\end{align}
\end{subequations}

In order to reduce the computational effort, the phase equilibrium calculations are performed uncoupled from the system of equations \ref{NR}. In these equations, when solving for \textbf{u}$^{t+1}$, all the fluid properties dependent on the phase equilibrium calculations ($x,y,\xi,g,\mathcal{L}, Z, H, \Delta P^{int}$) are computed with \textbf{u}$^{\tau}$, where $\tau$ corresponds to the preceding iteration of the Newton-Raphson method. In that way, instead of updating the phases saturations and properties only once per time-step, they are updated at every iteration between the solutions \textbf{u}$^{t+1}$ and \textbf{u}$^{t}$, improving the accuracy of the uncoupled solution.

\subsection{Phase Equilibrium Calculations}

For the phase equilibrium calculations, it is assumed constant temperature and local thermodynamic equilibrium at every time step within each pore. 
These calculations use a PT flash based on the \citet{Peng:1976} cubic equation of state. Therefore, at every time step of a simulation, once $P_i$ and $N_i^k$ are evaluated, a flash calculation is performed for each pore $i$ and new phase equilibrium conditions are determined. Details of the phase equilibrium calculations used in our model are described by \citet{santos}.

\section{Results}
\label{results}
\subsection{Validation}

 Relative permeability ($k_r$) curves obtained with coreflooding experiments by \citet{jamiolahmady2003} were used to validate the proposed model. The selected experiments were performed at three different gas flow rates and two different interfacial tension values, providing comprehensive data to evaluate whether the model is able to reproduce the effects of these parameters on gas and condensate coupled flow. Attempts to reproduce the same experiments with different pore-network models were presented by \citet{jamiolahmady2003} and \citet{momeni20173d}. Their results were also used in the validation and discussion.

\subsubsection{Pore-network Construction}
\begin{figure}[H]
    \centering
    \includegraphics[width=70mm]{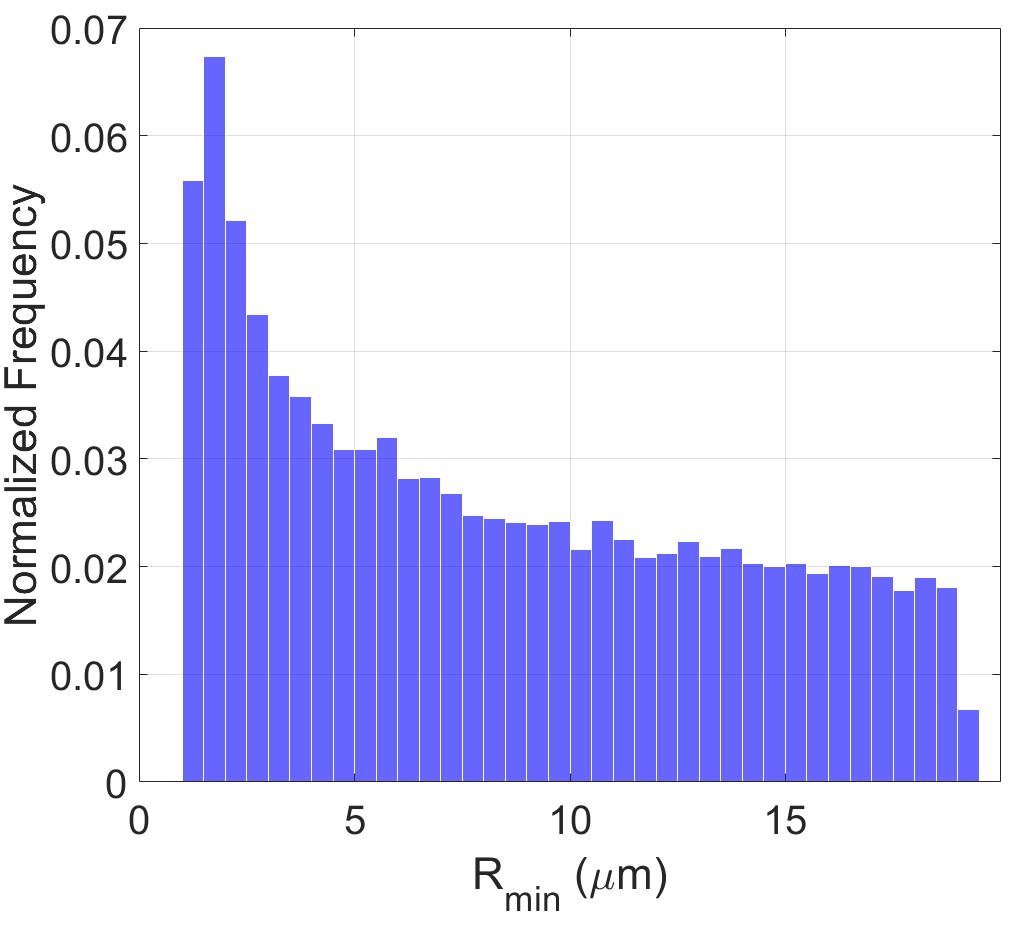}
    \caption{Normalized probability distribution function for $R_{min}$}
    \label{fig:Rmin}
\end{figure}

 A Berea core sample with porosity $\phi=18.2\%$, absolute permeability $k=92mD$ and irreducible water saturation $S_{wc}=26.4\%$ was used in the experiments \citet{jamiolahmady2003}. For simulation purposes, we disregarded the presence of water and gas/condensate saturations were corrected in the results to be based on  the total pore volume rather than the hydrocarbon pore volume. A Weibull probability distribution function of pore throat sizes was provided \cite{jamiolahmady2003} and used to define the constriction radius of the capillaries in the network. In this distribution, illustrated in Figure \ref{fig:Rmin}, $R_{min}$ ranges from $1.28\mu m$ to $19.21\mu m$, with an average value of $8.5\mu m$.
 
\begin{figure}[H]
    \centering
    \includegraphics[width=90mm]{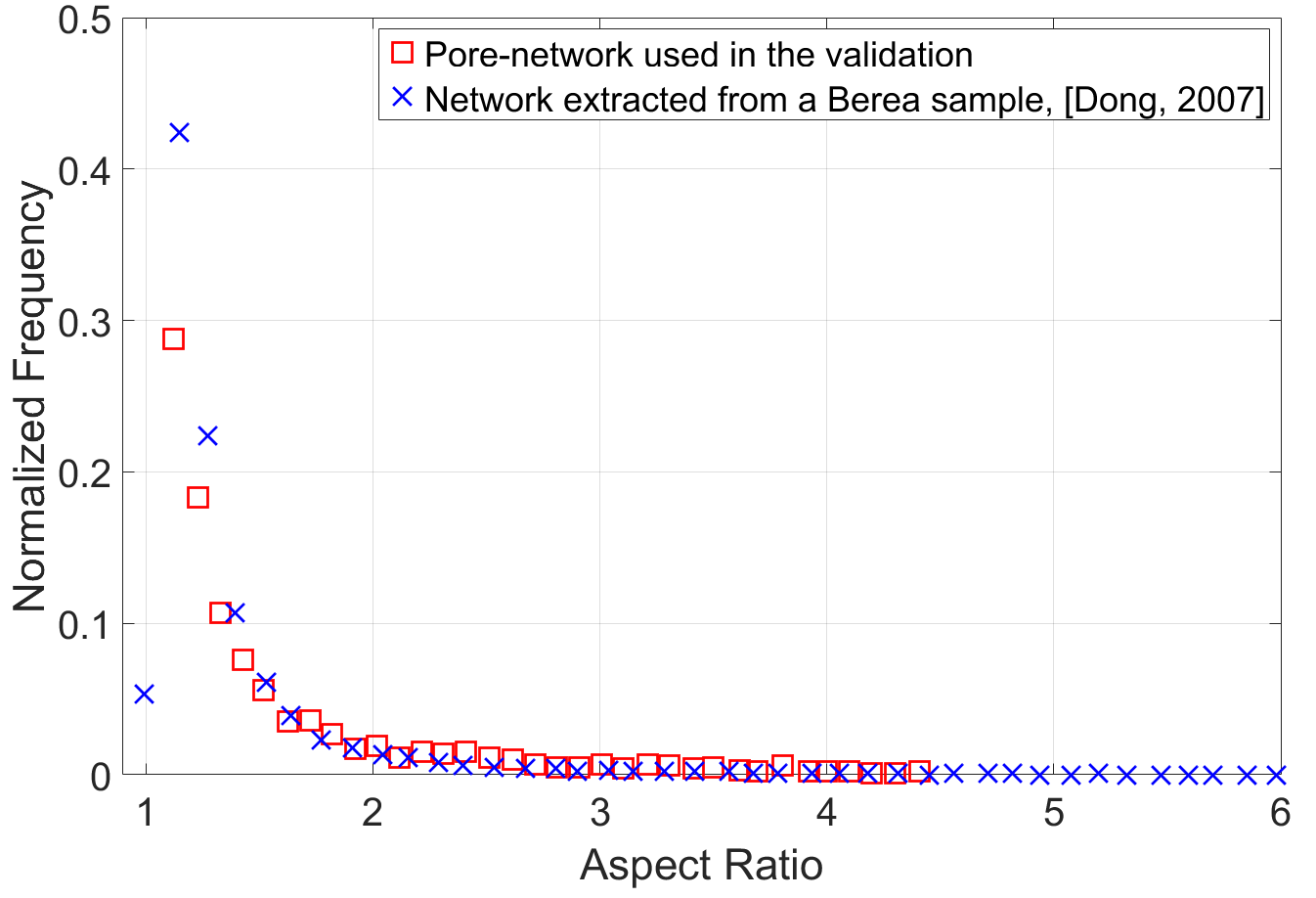}
    \caption{Aspect ratio of pore-networks extracted from a 3D image of a berea sandstone sample}
    \label{fig:AR}
\end{figure}

 In their model, \citet{jamiolahmady2003} used constricted capillaries with a constant ratio $R_{max}/R_{min}\approx3.33$. The authors suggested that this value did not reflect the true characteristics of the real porous medium and had been found to negatively affect the results. In the proposed model, this limitation was mitigated by calculating $R_{max}$ values based on data of the aspect ratio between pore body and throat radii, $AR=R_{max}/R_{min}$, of a berea sample. This correction is important since the aspect ratio has a large impact on multi-phase flow simulation \cite{dong2008micro}. The distribution function of the generated pore-network $AR$, along with that obtained for a pore-network extracted from a 3D Micro-CT image of a berea sample \cite{dong2008micro}, is presented in Figure \ref{fig:AR}, indicating that values of AR used in our model are representative of Berea cores. 
 
A cubic lattice of 20$\times$25$\times$25 nodes represented the 3D spatial distribution of pores in the network. The coordination number was set to have an average value of 3, and was obtained by randomly removing capillaries from the original cubic lattice, as done in the previous network models \cite{jamiolahmady2003,momeni20173d}. Additionally, the capillary length was chosen to be constant and equal to $75\mu m$. This value was found to satisfy the necessity of matching $\phi$ and $k$, with the set radii distribution, and also to be adequate to represent a berea sample \cite{dong2008micro}.

\subsubsection{Injected Fluid} 
Gas-condensate mixtures of methane ($C1$) and normal butane ($nC4$) were used as the fluid injected in the cores. The experiments were repeated at two levels of pressure, so that two different interfacial tension values were achieved. The pressures used in the simulations to reproduce the experimental conditions, for a fixed temperature of 37\textdegree C, as well as the viscosities of both phases calculated with the correlation proposed by \citet{lohrenz1964calculating}, are presented in Table \ref{data_validation}.

\begin{table}[h]
\centering
	\begin{tabular}{l l l l}	
			\hline
			\textbf{$\sigma$ (mN/m)} & \textbf{Pressure (MPa)}   & \textbf{$\mu_g$ (Pa.s)}   & \textbf{$\mu_l$ (Pa.s)}   \\
			\hline
			 0.015 & 13.04 & $2.26\times10^{-5}$ & $3.51\times10^{-5}$  \\
			 0.037 & 12.78 & $2.14\times10^{-5}$ & $3.71\times10^{-5}$  \\
			\hline
	\end{tabular}
	\caption{Pressure and viscosities of gas and condensate at two interfacial tension values and 37\textdegree C}
	\label{data_validation}      
\end{table}

\subsubsection{Gas and Condensate Flow Conditions}

Each point of the $k_r$ curves was measured by injecting the $C1-nC4$ mixture at a different condensate to gas flow ratio (CGR) in the cores. Also, for each CGR, the injected gas flow rate was controlled in order to generate gas flow velocities ($v_g)$ of $9md^{-1}$, $18md^{-1}$ and $36md^{-1}$. This was reproduced with the model by shifting the molar flow rate and molar fraction of the components in the injected composition. At a fixed pressure and temperature,  reducing the fraction of methane in the mixture, while increasing appropriately the total molar flow rate, leads to higher CGRs for the same $v_g$. For the case at $P_{out}=13.04MPa$, the $C1$ molar percentages were $75\%$, $77\%$, $78\%$, $79\%$ and $79.5\%$, while for the case at $P_{out}=12.78MPa$ they were $75\%$, $78\%$, $79\%$, $80\%$ and $80.5\%$. With these compositions, the initial condensate saturation in the networks ranged from $1\%$ to $22\%$ while the values achieved at steady-state flow varied from about $10\%$ to $40\%$. The criteria for achieving steady-state in the model, which was necessary for the relative permeability predictions, were equal injected and produced molar flow rates, and constant average pressure and saturation in the network.

\subsubsection{Relative Permeability Curves}

\begin{figure}[H]
    \centering
    \includegraphics[width=90mm]{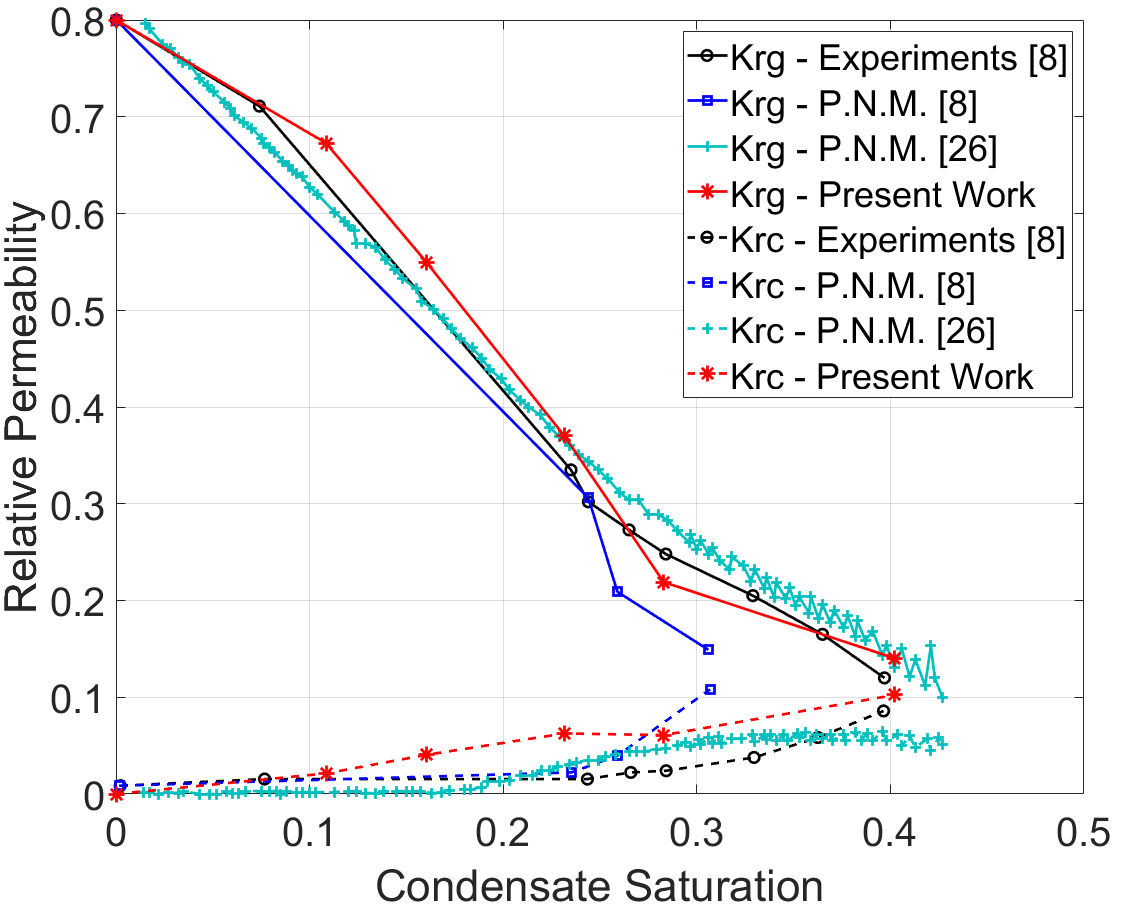}
    \caption{$k_{rg}$ and $k_{rc}$, at $\sigma=0.037mN/m$ and $v_g=18md^{-1}$}
    \label{fig:0037_18}
 \end{figure}

Figure \ref{fig:0037_18} presents the relative permeability curves for gas and condensate phases, $k_{rg}$ and $k_{rc}$, for $\sigma=0.037mN/m$ and $v_g=18md^{-1}$. It shows the predictions of the proposed model, curves obtained with experiments \cite{jamiolahmady2003} and curves predicted with the pore-network models presented by \citet{jamiolahmady2003} and \citet{momeni20173d}.

As discussed before, the model of \citet{jamiolahmady2003} does not consider the fluid composition and uses empirically adjusted coefficients to match the predictions to the measured relative permeabilities. Moreover, it assumes a fixed ratio between the pore body and pore throat radii. Although the model proposed by \citet{momeni20173d} uses a flash calculation to determine the amount of condensate at each throat, the system of equations is based on mass conservation in each pore and does not enforce molar conservation of each component. Also, in that work, the relative permeability curves are calculated in transient flow. At each time step, PT flash calculations are used to compute the volume of each phase in the throats and, consequently, the saturation in the network. The relative permeability of each phase is then calculated. This procedure leads to varying gas and condensate flow rates at the outlet of the network during the simulation and does not represent the experimental procedure adopted in \cite{jamiolahmady2003} to measure the relative permeabilities. 

All three models presented in the plot reproduce the experimental behavior of the gas relative permeability curve at low condensate saturation, $S_c<0.20$. While ours and Momeni et al.'s results follow the measured $k_{rg}$ curve up to $S_c\approx 0.4$, the predictions presented by \citet{jamiolahmady2003} present a sharp decline at $S_c\approx 0.25$, underpredicting the gas permeability for saturations higher than this value. Our model, while capturing the general behavior of the experimental $k_{rc}$ curve, overestimates the value, particularly in the range $0.15\leq S_c \leq 0.35$. This may be explained by the fact that the condensate conductance is calculated based on an annular flow pattern on smooth circular capillaries. Including the effect of wall roughness and non circular geometry can potentially reduce the mobility of the liquid phase and improve the results. Once again, the results from \citet{jamiolahmady2003} are accurate up to $S_c= 0.25$, above which they overpredict the experimentally measured $k_{rc}$. The curve presented by \citet{momeni20173d} does not capture the experimental curve behavior, with $k_{rc}\approx0$ for $S_c<0.17$ and a $k_{rc}$ plateau for $S_c>0.30$.

\begin{figure}[H]
    \centering
    \includegraphics[width=90mm]{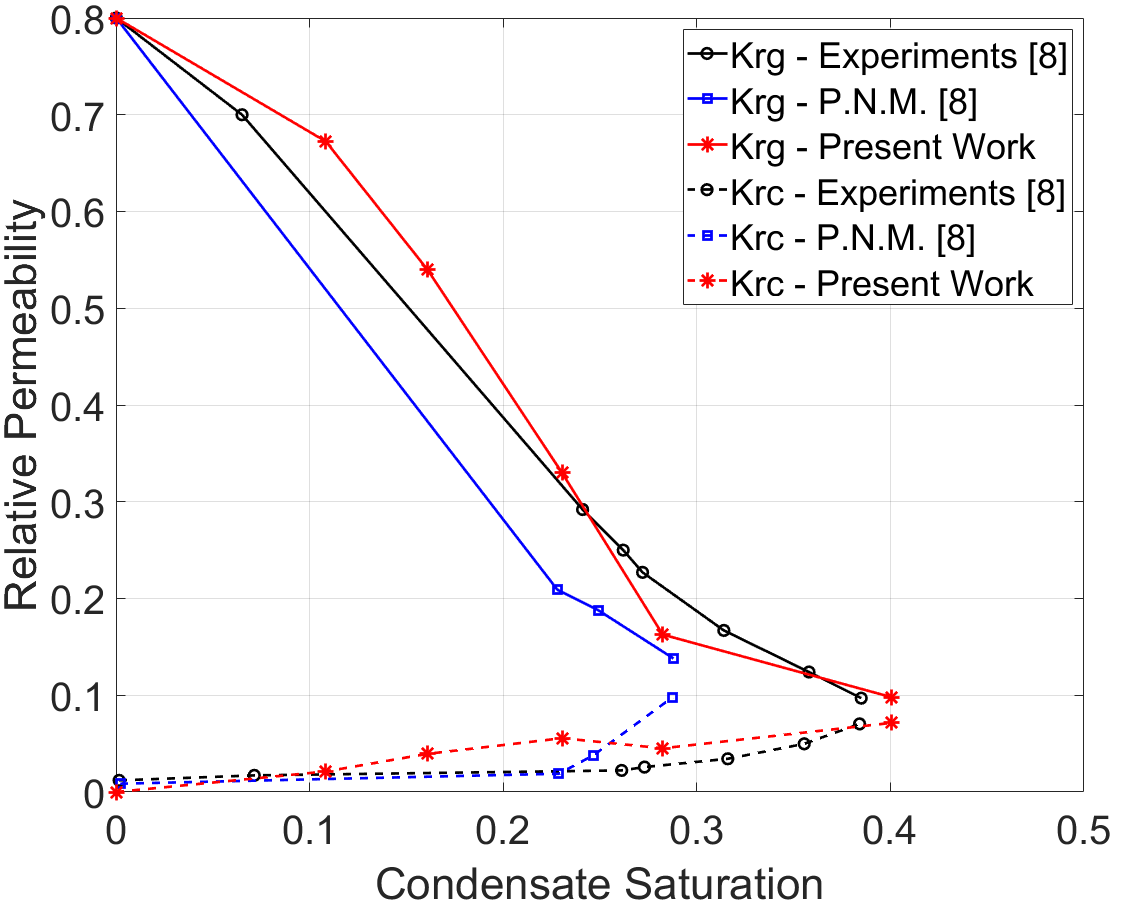}
    \caption{$k_{rg}$ and $k_{rc}$, at $\sigma=0.037mN/m$ and $v_g=9md^{-1}$}
    \label{fig:0037_9}
\end{figure}

\begin{figure}[H]
    \centering
    \includegraphics[width=90mm]{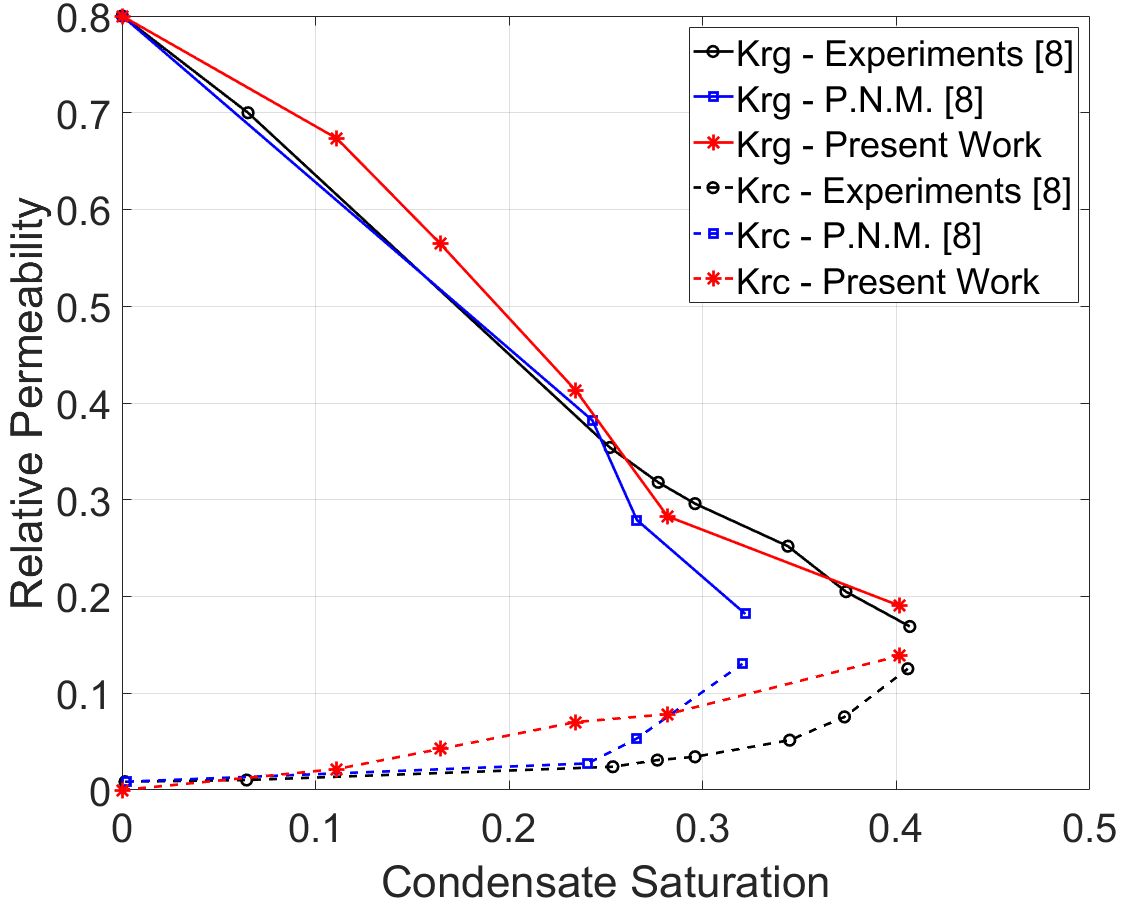}
    \caption{$k_{rg}$ and $k_{rc}$, at $\sigma=0.037mN/m$ and $v_g=36md^{-1}$}
    \label{fig:0037_36}
\end{figure}

The effect of gas velocity on the relative permeability curves is shown in  Figures \ref{fig:0037_9} and \ref{fig:0037_36}. They represent the $k_r$ curves for $\sigma =0.037mN/m$ and $v_g$ equal to $9md^{-1}$ and $36md^{-1}$, respectively. The proposed model is able to predict accurately the positive effect of flow rate on the relative permeability curves. For both gas velocities, our model produced results that more closely reproduce the experimental data then those presented by \citet{jamiolahmady2003}. \citet{momeni20173d} did not present results for $v_g$ other than $18md^{-1}$ or for different values of interfacial tension in their model validation.

\begin{figure}[H]
    \centering
    \includegraphics[width=90mm]{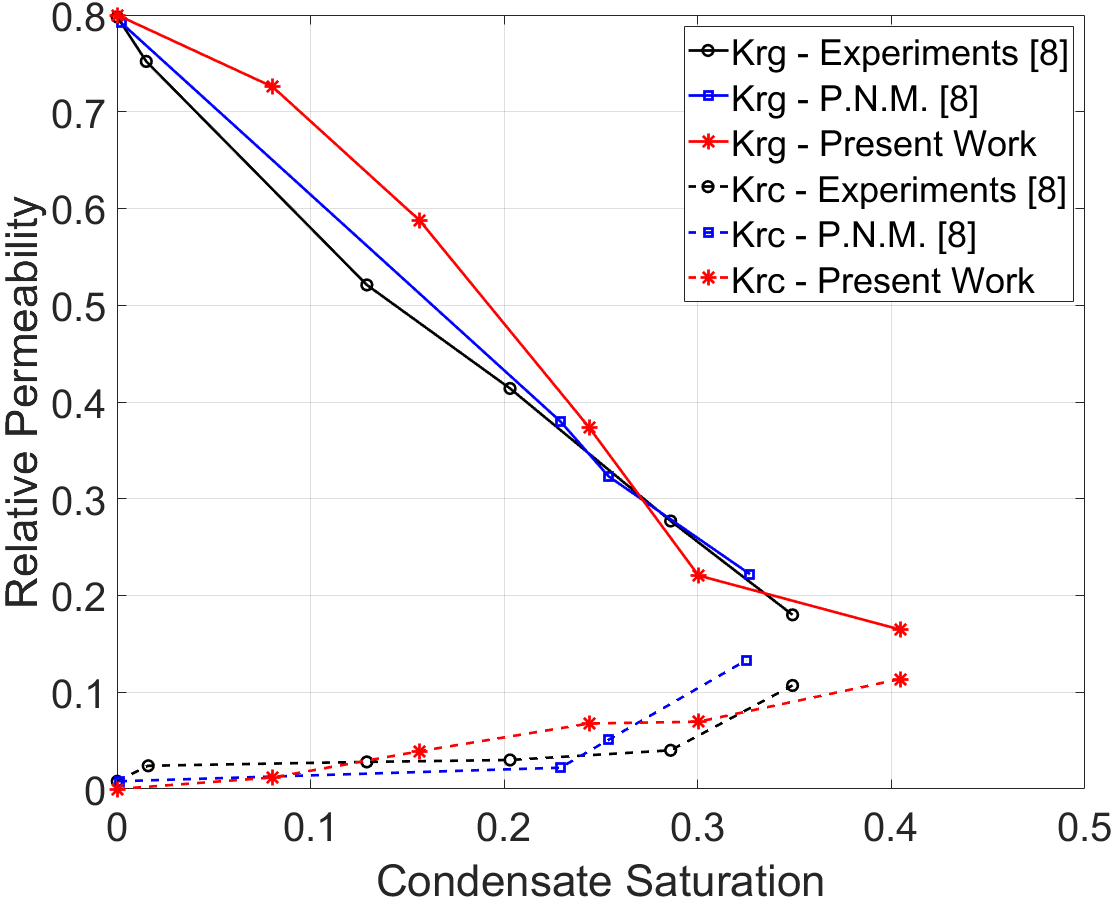}
    \caption{$k_{rg}$ and $k_{rc}$, at $\sigma=0.015mN/m$ and $v_g=9md^{-1}$}
    \label{fig:0015_9}
\end{figure}

\begin{figure}[H]
    \centering
    \includegraphics[width=90mm]{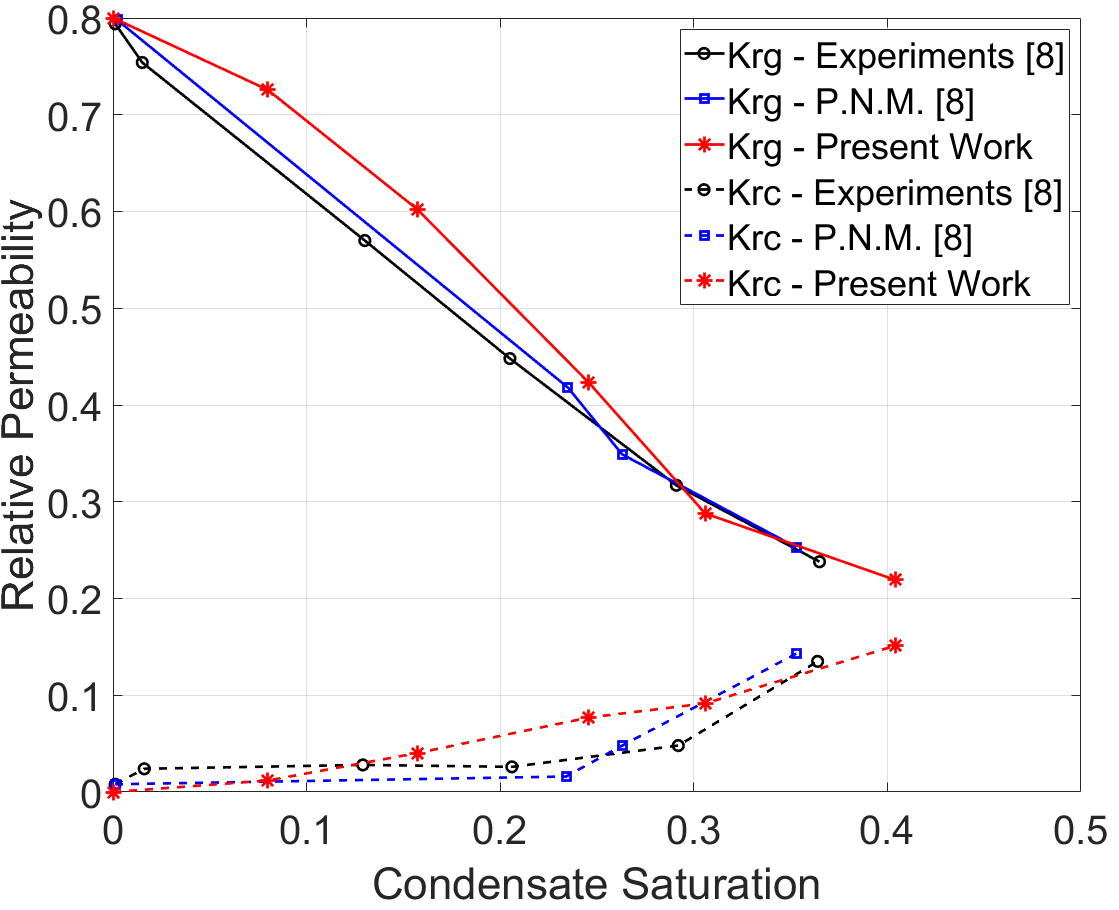}
    \caption{$k_{rg}$ and $k_{rc}$, at $\sigma=0.015mN/m$ and $v_g=18md^{-1}$}
    \label{fig:0015_18}
    
\end{figure}

\begin{figure}[H]
    \centering
    \includegraphics[width=90mm]{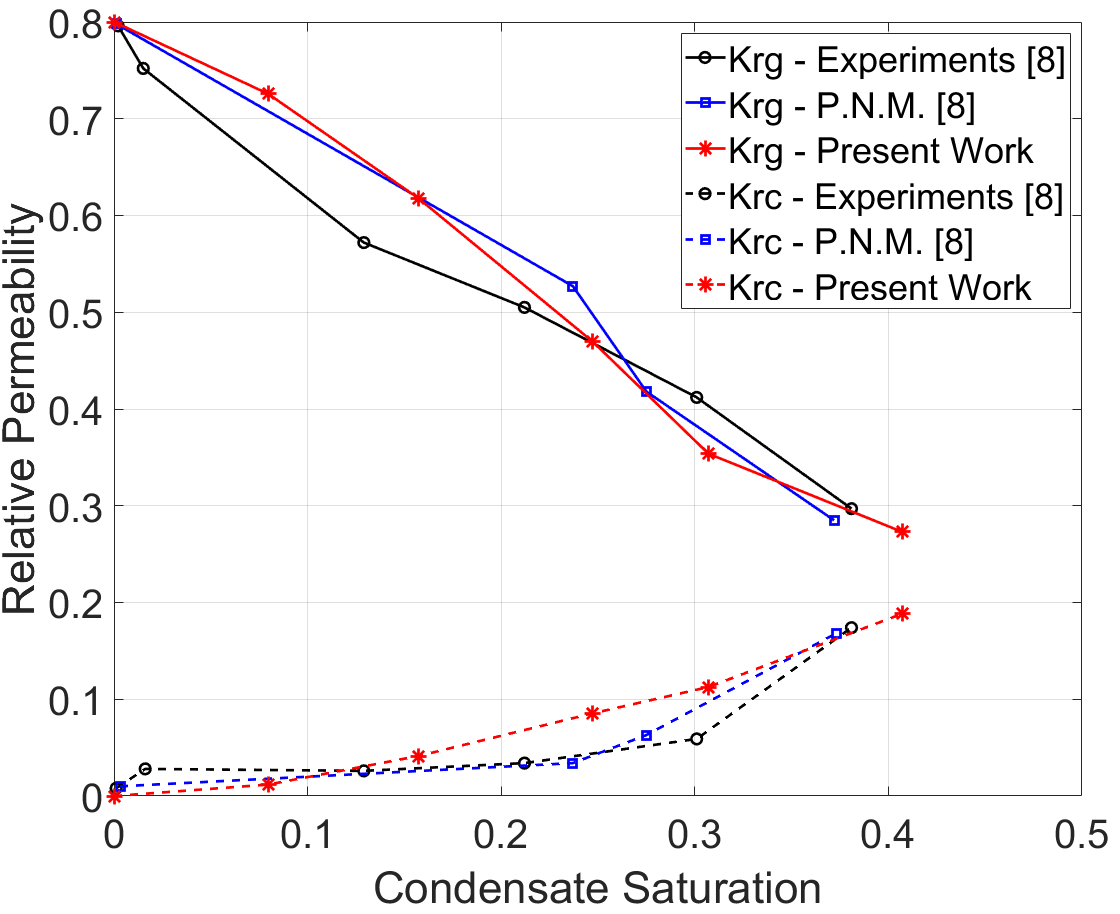}
    \caption{$k_{rg}$ and $k_{rc}$, at $\sigma=0.015mN/m$ and $v_g=36md^{-1}$}
    \label{fig:0015_36}
\end{figure}

The effect of interfacial tension on the flow behavior is presented in Figures \ref{fig:0015_9}, \ref{fig:0015_18} and \ref{fig:0015_36}, which show the relative permeability curves at low interfacial tension, $\sigma=0.015mN/m$, and different gas velocities. Higher $k_r$ are expected at lower interfacial tension levels, as the phase trapping mechanism is weakened with reduced capillary pressure. This effect was both verified experimentally and predicted with the proposed model. For the three analyzed gas flow velocities, the predicted $k_r$ curves agreed reasonably well with the measured data. While the model proposed by \citet{jamiolahmady2003} represented better the experimental $k_r$ curves at $v_g=9md^{-1}$ and $v_g=18md^{-1}$, their model had coefficients adjusted specifically to match the case at $\sigma=0.015mN/m$ and $v_g=9md^{-1}$, leading to the almost perfect fit. The proposed model does not have any adjustable parameter to fit experimental data.

\subsection{Visualization of Blocked Capillaries}

\begin{figure*}[t]
\begin{subfigure}{0.3\textwidth}
\includegraphics[width=1\linewidth]{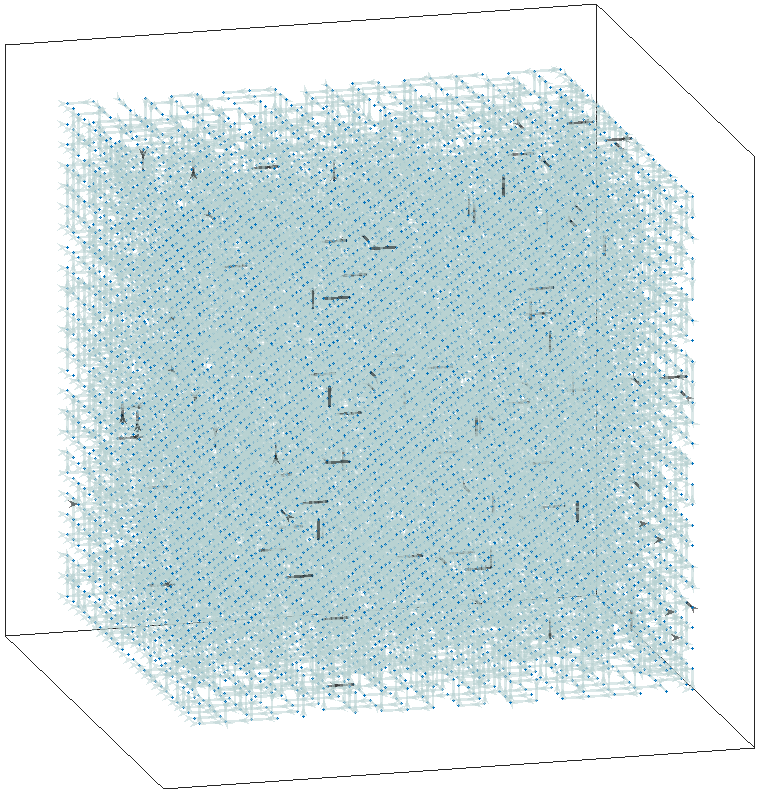}
\caption{$S_c=0.1083$} \label{fig:block_805}
\end{subfigure}
\hspace*{\fill} 
\begin{subfigure}{0.3\textwidth}
\includegraphics[width=1\linewidth]{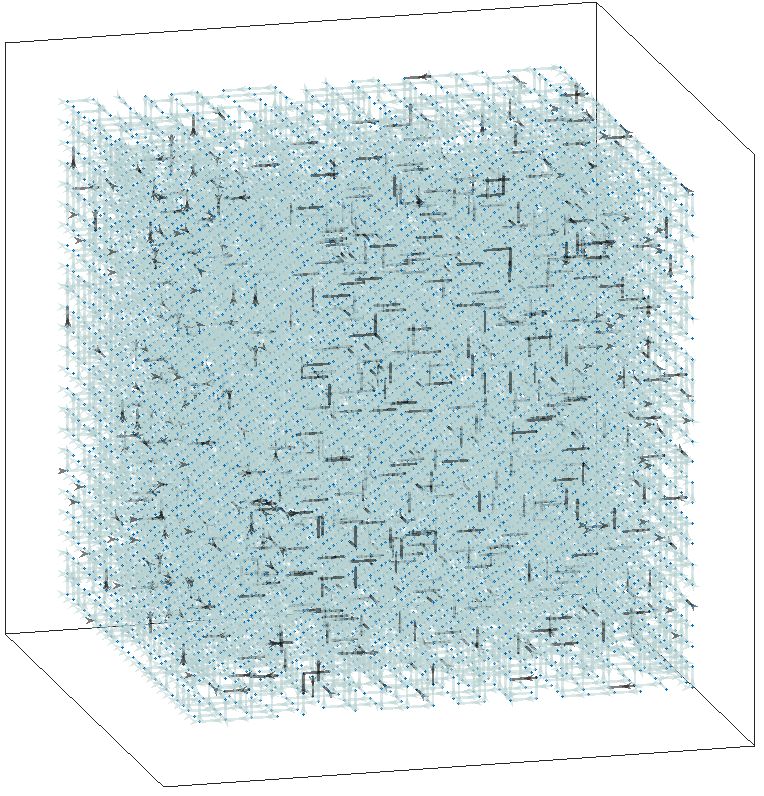}
\caption{$S_c=0.1609$} \label{fig:block_80}
\end{subfigure}
\hspace*{\fill} 
\begin{subfigure}{0.3\textwidth}
\includegraphics[width=1\linewidth]{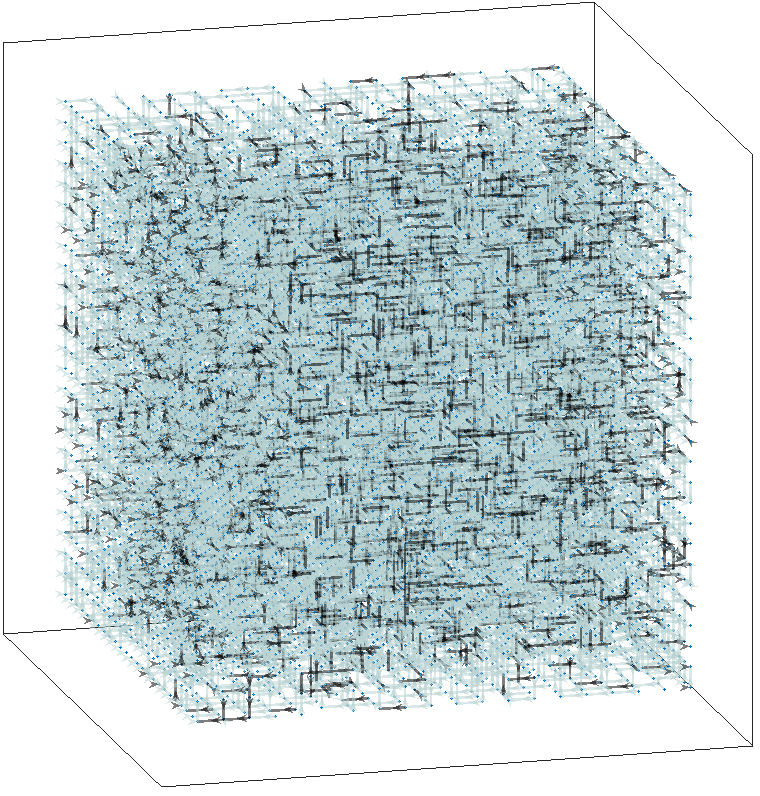}
\caption{$S_c=0.2304$} \label{fig:block_79}
\end{subfigure}
\hspace*{\fill} 
\begin{subfigure}{0.3\textwidth}
\includegraphics[width=1\linewidth]{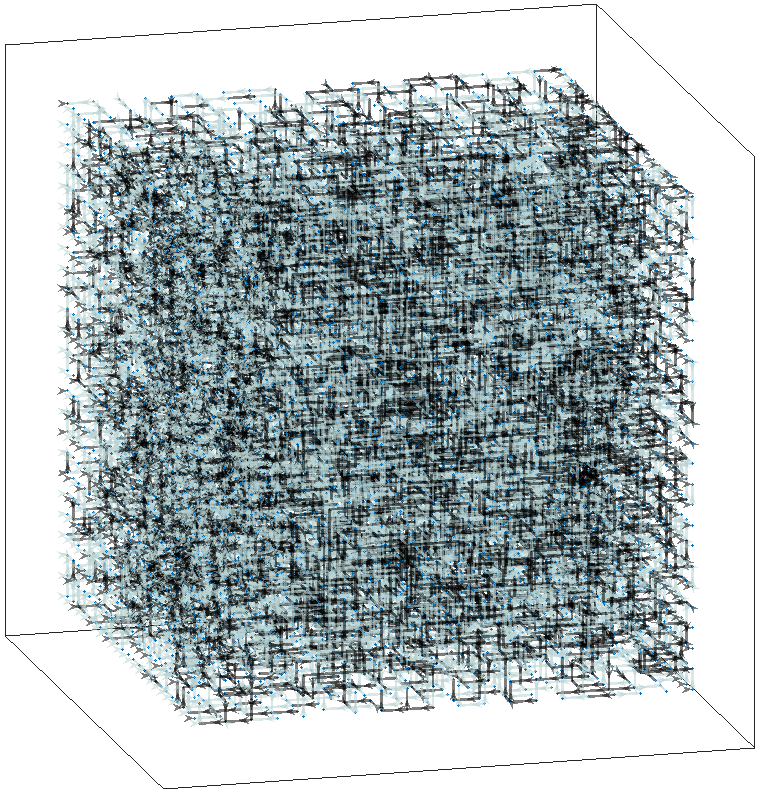}
\caption{$S_c=0.2824$} \label{fig:block_78}
\end{subfigure}
\hspace*{\fill} 
\begin{subfigure}{0.3\textwidth}
\includegraphics[width=1\linewidth]{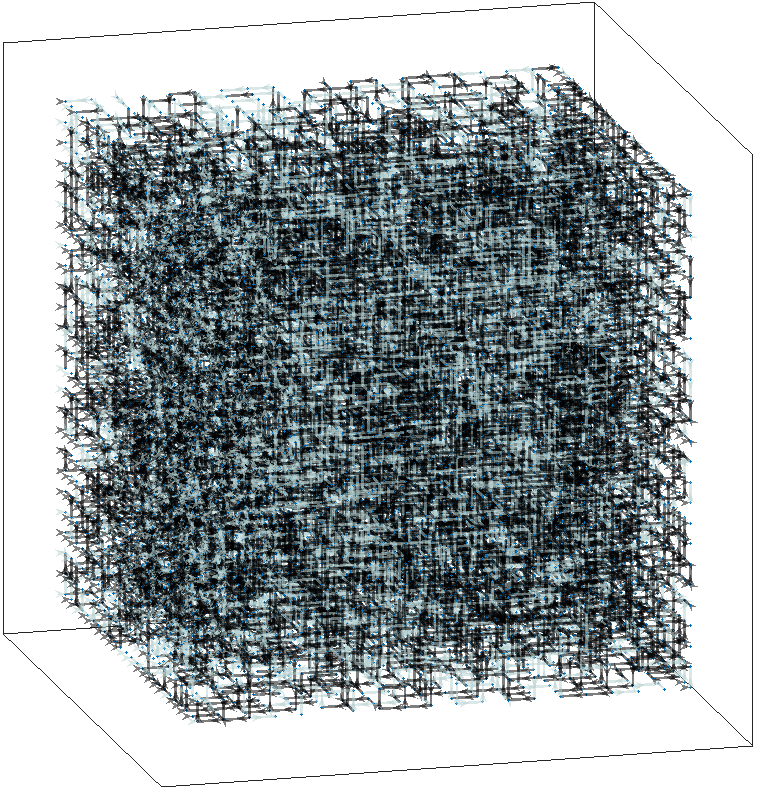}
\caption{$S_c=0.4007$} \label{fig:block_75}
\end{subfigure}
\begin{subfigure}{0.2\textwidth}
\includegraphics[width=1\linewidth]{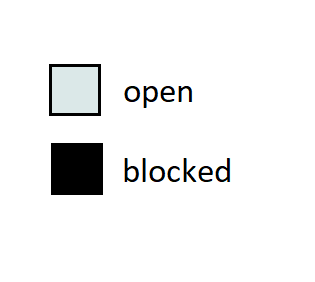}
\end{subfigure}
\hspace*{\fill} 

\caption{Blocked Capillaries, at $\sigma=0.037mN/m$ and $v_g=18md^{-1}$} 
\label{fig:block_cap}

\end{figure*}

The sharp decline in $k_{rg}$ with condensate saturation buildup in porous media has been attributed to the formation of condensate bridges that obstruct the gas flow in pore throats \cite{Harassi2009,Coskuner1997,jamiolahmady2000,jamiolahmady2003}. This phenomenon was well captured by our model, as illustrated in Figure \ref{fig:block_cap}. Figures \ref{fig:block_cap} (a) to (e) show the opened and blocked capillaries of the network at the conditions used to construct the $k_r$ curves at $\sigma=0.037mN/m$ and $v_g=18md^{-1}$, displayed in Figure \ref{fig:0037_18}. For the two lowest $S_c$ values, only a small fraction of capillaries is obstructed with condensate, $0.5\%$ and $3.64\%$, for (a) and (b) respectively, meaning that $k_{rg}$ is not significantly lessened. Also, at those saturation levels, the blocked capillaries not only are not numerous, but also are the most constricted ones, which implies that they provided a weak contribution to flow. For the three highest $S_c$ values, the percentage of blocked capillaries rose to (c) $14.36\%$, (d) $43.50\%$ and (e) $64.73\%$, which reduces significantly the available gas flowing paths, leading to a substantial decrease in $k_{rg}$

\subsection{Compositional Shift in Gas Condensate Flow}

A compositional model based on molar conservation of each fluid component and phase equilibrium calculations can be used to study the changes in composition of each phase as they flow through a porous medium. As the pressure in the network is lowered below the $P_{dp}$, the liquid phase that emerges is richer in heavier hydrocarbon components than the remaining gas phase. Since the flow of gas and liquid in small microchannels presents a large slip ratio, meaning that the gas velocity is higher than the condensate velocity \cite{chung2004effect,kawahara2002investigation,kawaji2004adiabatic}, this leads to an accumulation of liquid and, hence, heavy components in the porous media. This can be exemplified in Table \ref{comp_shift}. It contains the molar percentage of n-butane in the fluid injected in the network, and the molar percentage of the same component found in the network at steady-state condition, at $\sigma=0.037mN/m$ and $v_g=18md^{-1}$.

\begin{table}[h]
\centering
\begin{tabular}{l l}	
\hline
\textbf{Injected $nC4\%$} &  \textbf{$nC4\%$ in the Network}  \\
\hline
 19.5 & 22.14  \\
 20.0 & 23.51  \\
 21.0 & 25.29 \\
 22.0 & 26.52 \\
 25.0 & 29.19 \\
 \hline
\end{tabular}
\caption{Compositional shift in the network at at $\sigma=0.037mN/m$ and $v_g=18md^{-1}$}
\label{comp_shift}       
\end{table}

\begin{figure}[H]
    \centering
    \includegraphics[width=90mm]{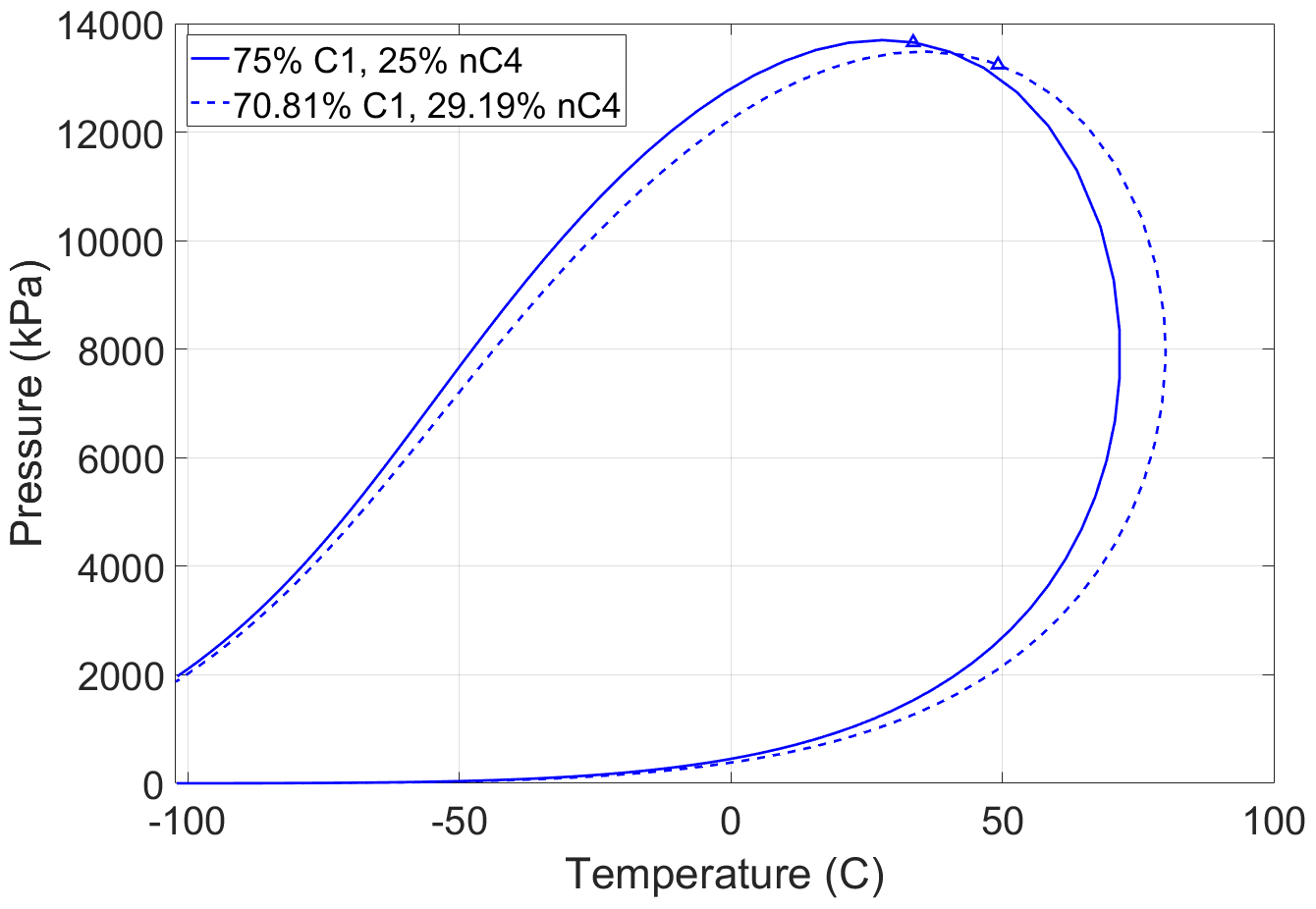}
    \caption{Phase Envelopes of a C1-nC4 binary mixture}
    \label{fig:PE_75}
\end{figure}

For all injection scenarios, the fraction of $nC4$, the heavier component in the studied binary mixture, was higher in the network than in the injected fluid. Consequently, the $C1-nC4$ phase envelopes are displaced to the right, as illustrated in Figure \ref{fig:PE_75}, for the case where the $nC4$ raised from $25\%$ to $29.19\%$. This shift not only impacts the liquid dropout prediction, but also may lead to a transition in the general behavior of the mixture. It can be seen in the phase envelopes that the critical point, marked as a  $\bigtriangleup$, was dislocated from T=33\textdegree C to T=49\textdegree C. Given that the network temperature was 37\textdegree C, this makes the behavior of the fluid mixture change from gas-condensate to volatile oil. This phenomenon, known to take place in gas condensate reservoirs near the wellbore region \cite{roussenac2001}, impacts significantly the response of the fluid system to changes in pressure and temperature, and can only be appropriately predicted with compositional modeling, such as the one proposed here. 

\section{Conclusions}

A new compositional dynamic pore-network model was presented to study gas and condensate coupled flow in porous media. The network comprised a 3D structure of pores connected by constricted circular capillaries with hyperbolic profile. Gas and condensate conductances in the capillaries were calculated based on annular flow pattern, with liquid wetting capillary walls. Interruption of annular flow in the capillaries by the formation of condensate bridges was predicted, based on geometry and local liquid saturation. Once blocked, a capillary could be reopened if the pressure difference between its extremities exceeded a critical capillary pressure. Molar content and pressure in each pore were calculated through the solution of a system of non-linear equations with Newton-Raphson method, while the phases saturations were determined with a PT flash.

The model was validated by comparing predictions of steady-state gas and condensate relative permeability curves with experimental data available in the literature. Results demonstrated that the model represented well the effect of condensate saturation, interfacial tension and velocity on gas-condensate flow. Predicted $k_r$ curves for flow at two different interfacial tension values and three different gas velocities demonstrated reasonable quantitative agreement with experimental data. Also, analyses of the fraction of blocked capillaries in the networks, for different condensate saturations, indicated that the formation and retention of condensate slugs in pore throats could constitute the main mechanism related to condensate banking. 

The model also proved to be capable of predicting the buildup of heavy components in porous media during gas and condensate flow. For the $C1-nC4$ binary mixture used in the study, at all tested injection conditions, an increase of n-butane molar fraction was verified in the network. For the cases at high $S_c$, a transition in behavior from gas-condensate to volatile oil was observed.

With the provided evidence of predictive capability, we believe that the proposed model has potential to produce tailored data to large-scale gas-condensate reservoir simulation, based on specific reservoir rock, fluid composition, and flow conditions. This could lead to more realistic production estimations and benefit gas-condensate fields development planning. Future work in the model could encompass the implementation of wall roughness effect in condensate flow and the construction of more realistic pore-networks, generated based on pore-scale imaging techniques.

\section{Acknowlegments}

This research was funded by Repsol-Sinopec Brasil, under the RD\&I Levy Fund Program of the National Petroleum Agency (ANP).


\appendix
\section{Functions $H$ and $H^{int}$}
\label{app:H}
 
Predicting whether a capillary is opened or closed to flow, due to the formation and retention of condensate bridges, constitutes a central part in the proposed model formulation. The flow status of a capillary, as available or not for flow, can be written as a function of its gas saturation and pressure difference. For relatively high $S_g$ and/or high $\Delta P$, the capillary is opened to flow. The combination of low $S_g$ and $\Delta P$ leads to a capillary that is closed to flow. Including this transitions in the model would, therefore, make the conductances discontinuous function of $S_g$ and $\Delta P$. Another instance of discontinuity in the model formulation involves the inclusion the interfacial pressure difference in the calculation of the capillary flow rate. This term only appears for low $S_g$ and high $\Delta P$, meaning that a condensate slug is moving through the capillary.

However, as a condition for convergence of the multivariate Newton-Raphson method, the system of equations should contain only continuously differentiable functions in the neighborhood of its roots. Therefore, in order to avoid convergence issues, those discontinuous functions of $S_g$ and $\Delta P$ were substituted by the continuous approximations of unit step functions given by Equation \ref{h1h2}.

\begin{subequations}
\label{h1h2}
\begin{align}
 h_1= \frac{1}{2} \left\{ 1 - \tanh{ \left[\frac{1}{K_1} \left( \frac{S_g}{S_{g,crit}}-1\right) \right] } \right\} \\
 h_2= \frac{1}{2} \left\{ 1 - \tanh{ \left[\frac{1}{K_2} \left( \frac{\Delta P}{\Delta P_{crit}}-1\right) \right] } \right\}
 \end{align}
\end{subequations}

In Equation \ref{h1h2},  $S_{g,crit}$ is the gas saturation  calculated at a condensate film thickness $t=t_{crit}$. Thus, $S_g<S_{g,crit}$ leads to the occurrence of snap-off. 
Function $h_1$ is equal to one for $S_g < S_{g,crit} - \epsilon$ and equal to zero for 
$ S_g > S_{g,crit} + \epsilon$. $2 \epsilon$ is the thickness of the continuous representation region of the discontinuity of the step function. $\Delta P_{crit}$ is the critical capillary pressure difference for a capillary to be reopened to flow, once the snap-off happened, given by eq. \ref{dP_crit}. Function $h_2$ is equal to one for $\Delta P<\Delta P_{crit}- \epsilon$ and equal to zero for $\Delta P>\Delta P_{crit}+ \epsilon$.
$K_1$ and $K_2$ are shape factors that control the thickness $2 \epsilon$ on
the transition region. Low values for $K_1$ and $K_2$ lead to smaller $\epsilon$ and
better approximations of a discrete step function, and their values should be adjusted according to functions $h_1$ and $h_2$ domains.

Since the capillary flow is interrupted when both $S_g$ and $\Delta P$ are below their critical values, the conductances were multiplied by $H$, given by equation \ref{eq:H}, in the capillary molar flow calculation (eq. \ref{nj}).

\begin{equation}
\label{eq:H}
 H = 1-h_1h_2 
\end{equation}

As for the subtraction of the interfacial pressure drop in the capillary flow rate calculation, it was achieved by multiplying $\Delta P ^{int}$ by $H^{int}$, shown in Equation \ref{eq:H}. 

\begin{equation}
\label{eq:Hint}
 H^{int} = h_1-h_1h_2 
\end{equation}

For the results presented in section \ref{results}, the adopted shape factors were $K_1=0.1$ and $K_2=0.3$. This values were achieved upon successive trials of reducing the shape factors, and proved to be the low enough to represent well a unit step function, but not so low as to cause convergence difficulties.




\bibliographystyle{elsarticle-harv}\biboptions{authoryear}
\bibliography{PNMGC}

\begin{thebibliography}{37}
\expandafter\ifx\csname natexlab\endcsname\relax\def\natexlab#1{#1}\fi
\providecommand{\url}[1]{\texttt{#1}}
\providecommand{\href}[2]{#2}
\providecommand{\path}[1]{#1}
\providecommand{\DOIprefix}{doi:}
\providecommand{\ArXivprefix}{arXiv:}
\providecommand{\URLprefix}{URL: }
\providecommand{\Pubmedprefix}{pmid:}
\providecommand{\doi}[1]{\href{http://dx.doi.org/#1}{\path{#1}}}
\providecommand{\Pubmed}[1]{\href{pmid:#1}{\path{#1}}}
\providecommand{\bibinfo}[2]{#2}
\ifx\xfnm\relax \def\xfnm[#1]{\unskip,\space#1}\fi
\bibitem[{Al~Harrasi et~al.(2009)Al~Harrasi, Grattoni, Fisher and
  Al-Hinai}]{Harassi2009}
\bibinfo{author}{Al~Harrasi, M.}, \bibinfo{author}{Grattoni, C.},
  \bibinfo{author}{Fisher, Q.}, \bibinfo{author}{Al-Hinai, S.},
  \bibinfo{year}{2009}.
\newblock \bibinfo{title}{Condensate displacement mechanisms in low
  permeability rocks}, in: \bibinfo{booktitle}{Proceedings},
  \bibinfo{organization}{Society of Core Analysts}. p.~\bibinfo{pages}{15}.
\bibitem[{Azin et~al.(2019)Azin, Sedaghati, Fatehi, Osfouri and
  Sakhaei}]{azin2019production}
\bibinfo{author}{Azin, R.}, \bibinfo{author}{Sedaghati, H.},
  \bibinfo{author}{Fatehi, R.}, \bibinfo{author}{Osfouri, S.},
  \bibinfo{author}{Sakhaei, Z.}, \bibinfo{year}{2019}.
\newblock \bibinfo{title}{Production assessment of low production rate of well
  in a supergiant gas condensate reservoir: application of an integrated
  strategy}.
\newblock \bibinfo{journal}{Journal of Petroleum Exploration and Production
  Technology} \bibinfo{volume}{9}, \bibinfo{pages}{543--560}.
\bibitem[{Barnum et~al.(1995)Barnum, Brinkman, Richardson, Spillette
  et~al.}]{barnum1995gas}
\bibinfo{author}{Barnum, R.}, \bibinfo{author}{Brinkman, F.},
  \bibinfo{author}{Richardson, T.}, \bibinfo{author}{Spillette, A.}, et~al.,
  \bibinfo{year}{1995}.
\newblock \bibinfo{title}{Gas condensate reservoir behaviour: productivity and
  recovery reduction due to condensation}, in: \bibinfo{booktitle}{SPE annual
  technical conference and exhibition}, \bibinfo{organization}{Society of
  Petroleum Engineers}.
\bibitem[{Beresnev et~al.(2009)Beresnev, Li and Vigil}]{beresnev2009condition}
\bibinfo{author}{Beresnev, I.A.}, \bibinfo{author}{Li, W.},
  \bibinfo{author}{Vigil, R.D.}, \bibinfo{year}{2009}.
\newblock \bibinfo{title}{Condition for break-up of non-wetting fluids in
  sinusoidally constricted capillary channels}.
\newblock \bibinfo{journal}{Transport in porous media} \bibinfo{volume}{80},
  \bibinfo{pages}{581}.
\bibitem[{Blunt et~al.(2013)Blunt, Bijeljic, Dong, Gharbi, Iglauer, Mostaghimi,
  Paluszny and Pentland}]{blunt2013pore}
\bibinfo{author}{Blunt, M.J.}, \bibinfo{author}{Bijeljic, B.},
  \bibinfo{author}{Dong, H.}, \bibinfo{author}{Gharbi, O.},
  \bibinfo{author}{Iglauer, S.}, \bibinfo{author}{Mostaghimi, P.},
  \bibinfo{author}{Paluszny, A.}, \bibinfo{author}{Pentland, C.},
  \bibinfo{year}{2013}.
\newblock \bibinfo{title}{Pore-scale imaging and modelling}.
\newblock \bibinfo{journal}{Advances in Water resources} \bibinfo{volume}{51},
  \bibinfo{pages}{197--216}.
\bibitem[{Bustos and Toledo(2003)}]{bustos2003}
\bibinfo{author}{Bustos, C.I.}, \bibinfo{author}{Toledo, P.G.},
  \bibinfo{year}{2003}.
\newblock \bibinfo{title}{Pore-level modeling of gas and condensate flow in
  two-and three-dimensional pore networks: Pore size distribution effects on
  the relative permeability of gas and condensate}.
\newblock \bibinfo{journal}{Transport in Porous Media} \bibinfo{volume}{53},
  \bibinfo{pages}{281--315}.
\bibitem[{Chen et~al.(1995)Chen, Wilson and Monger-McClure}]{Chen1995}
\bibinfo{author}{Chen, H.L.}, \bibinfo{author}{Wilson, S.D.},
  \bibinfo{author}{Monger-McClure, T.G.}, \bibinfo{year}{1995}.
\newblock \bibinfo{title}{Determination of relative permeability and recovery
  for north sea gas condensate reservoirs}, in: \bibinfo{booktitle}{SPE Annual
  Technical Conference and Exhibition}, \bibinfo{organization}{Society of
  Petroleum Engineers}.
\bibitem[{Chung and Kawaji(2004)}]{chung2004effect}
\bibinfo{author}{Chung, P.Y.}, \bibinfo{author}{Kawaji, M.},
  \bibinfo{year}{2004}.
\newblock \bibinfo{title}{The effect of channel diameter on adiabatic two-phase
  flow characteristics in microchannels}.
\newblock \bibinfo{journal}{International journal of multiphase flow}
  \bibinfo{volume}{30}, \bibinfo{pages}{735--761}.
\bibitem[{Co{\c{s}}kuner(1997)}]{Coskuner1997}
\bibinfo{author}{Co{\c{s}}kuner, G.}, \bibinfo{year}{1997}.
\newblock \bibinfo{title}{Microvisual study of multiphase gas condensate flow
  in porous media}.
\newblock \bibinfo{journal}{Transport in Porous Media} \bibinfo{volume}{28},
  \bibinfo{pages}{1--18}.
\newblock \DOIprefix\doi{10.1023/A:1006505706431}.
\bibitem[{Davarpanah et~al.(2019)Davarpanah, Mazarei and
  Mirshekari}]{davarpanah2019simulation}
\bibinfo{author}{Davarpanah, A.}, \bibinfo{author}{Mazarei, M.},
  \bibinfo{author}{Mirshekari, B.}, \bibinfo{year}{2019}.
\newblock \bibinfo{title}{A simulation study to enhance the gas production rate
  by nitrogen replacement in the underground gas storage performance}.
\newblock \bibinfo{journal}{Energy Reports} \bibinfo{volume}{5},
  \bibinfo{pages}{431--435}.
\bibitem[{Dawe and Grattoni(2007)}]{Dawe2007}
\bibinfo{author}{Dawe, R.A.}, \bibinfo{author}{Grattoni, C.A.},
  \bibinfo{year}{2007}.
\newblock \bibinfo{title}{{Fluid flow behaviour of gas-condensate and
  near-miscible fluids at the pore scale}}.
\newblock \bibinfo{journal}{Journal of Petroleum Science and Engineering}
  \bibinfo{volume}{55}, \bibinfo{pages}{228--236}.
\newblock \DOIprefix\doi{10.1016/j.petrol.2006.08.009}.
\bibitem[{Dong(2008)}]{dong2008micro}
\bibinfo{author}{Dong, H.}, \bibinfo{year}{2008}.
\newblock \bibinfo{title}{Micro-CT imaging and pore network extraction}.
\newblock Ph.D. thesis. Department of Earth Science and Engineering, Imperial
  College London.
\bibitem[{El-Banbi et~al.(2000)El-Banbi, McCain~Jr, Semmelbeck
  et~al.}]{Banbi2000}
\bibinfo{author}{El-Banbi, A.H.}, \bibinfo{author}{McCain~Jr, W.},
  \bibinfo{author}{Semmelbeck, M.}, et~al., \bibinfo{year}{2000}.
\newblock \bibinfo{title}{Investigation of well productivity in gas-condensate
  reservoirs}, in: \bibinfo{booktitle}{SPE/CERI Gas Technology Symposium},
  \bibinfo{organization}{Society of Petroleum Engineers}.
\bibitem[{Fang et~al.(1996)Fang, Firoozabadi, Abbaszadeh, Radke
  et~al.}]{fang1996}
\bibinfo{author}{Fang, F.}, \bibinfo{author}{Firoozabadi, A.},
  \bibinfo{author}{Abbaszadeh, M.}, \bibinfo{author}{Radke, C.}, et~al.,
  \bibinfo{year}{1996}.
\newblock \bibinfo{title}{A phenomenological modeling of critical condensate
  saturation}, in: \bibinfo{booktitle}{SPE Annual Technical Conference and
  Exhibition}, \bibinfo{organization}{Society of Petroleum Engineers}.
\bibitem[{Henderson et~al.(1998)Henderson, Danesh, Tehrani, Al-Shaidi, Peden
  et~al.}]{kr_rate_2}
\bibinfo{author}{Henderson, G.D.}, \bibinfo{author}{Danesh, A.},
  \bibinfo{author}{Tehrani, D.H.}, \bibinfo{author}{Al-Shaidi, S.},
  \bibinfo{author}{Peden, J.M.}, et~al., \bibinfo{year}{1998}.
\newblock \bibinfo{title}{Measurement and correlation of gas condensate
  relative permeability by the steady-state method}.
\newblock \bibinfo{journal}{SPE Reservoir Evaluation \& Engineering}
  \bibinfo{volume}{1}, \bibinfo{pages}{134--140}.
\bibitem[{Hoseinpour et~al.(2019)Hoseinpour, Madhi, Norouzi, Soulgani and
  Mohammadi}]{hoseinpour2019condensate}
\bibinfo{author}{Hoseinpour, S.A.}, \bibinfo{author}{Madhi, M.},
  \bibinfo{author}{Norouzi, H.}, \bibinfo{author}{Soulgani, B.S.},
  \bibinfo{author}{Mohammadi, A.H.}, \bibinfo{year}{2019}.
\newblock \bibinfo{title}{Condensate blockage alleviation around gas-condensate
  producing wells using wettability alteration}.
\newblock \bibinfo{journal}{Journal of Natural Gas Science and Engineering}
  \bibinfo{volume}{62}, \bibinfo{pages}{214--223}.
\bibitem[{Jamiolahmady et~al.(2003a)Jamiolahmady, Danesh, Henderson, Tehrani
  et~al.}]{kr_rate_1}
\bibinfo{author}{Jamiolahmady, M.}, \bibinfo{author}{Danesh, A.},
  \bibinfo{author}{Henderson, G.}, \bibinfo{author}{Tehrani, D.}, et~al.,
  \bibinfo{year}{2003}a.
\newblock \bibinfo{title}{Variations of gas-condensate relative permeability
  with production rate at near wellbore conditions: a general correlation}, in:
  \bibinfo{booktitle}{Offshore Europe}, \bibinfo{organization}{Society of
  Petroleum Engineers}.
\bibitem[{Jamiolahmady et~al.(2000)Jamiolahmady, Danesh, Tehrani and
  Duncan}]{jamiolahmady2000}
\bibinfo{author}{Jamiolahmady, M.}, \bibinfo{author}{Danesh, A.},
  \bibinfo{author}{Tehrani, D.H.}, \bibinfo{author}{Duncan, D.B.},
  \bibinfo{year}{2000}.
\newblock \bibinfo{title}{A mechanistic model of gas-condensate flow in pores}.
\newblock \bibinfo{journal}{Transport in porous media} \bibinfo{volume}{41},
  \bibinfo{pages}{17--46}.
\bibitem[{Jamiolahmady et~al.(2003b)Jamiolahmady, Danesh, Tehrani and
  Duncan}]{jamiolahmady2003}
\bibinfo{author}{Jamiolahmady, M.}, \bibinfo{author}{Danesh, A.},
  \bibinfo{author}{Tehrani, D.H.}, \bibinfo{author}{Duncan, D.B.},
  \bibinfo{year}{2003}b.
\newblock \bibinfo{title}{Positive effect of flow velocity on gas--condensate
  relative permeability: network modelling and comparison with experimental
  results}.
\newblock \bibinfo{journal}{Transport in Porous Media} \bibinfo{volume}{52},
  \bibinfo{pages}{159--183}.
\bibitem[{Joekar-Niasar and Hassanizadeh(2012)}]{joekar2012analysis}
\bibinfo{author}{Joekar-Niasar, V.}, \bibinfo{author}{Hassanizadeh, S.},
  \bibinfo{year}{2012}.
\newblock \bibinfo{title}{Analysis of fundamentals of two-phase flow in porous
  media using dynamic pore-network models: A review}.
\newblock \bibinfo{journal}{Critical reviews in environmental science and
  technology} \bibinfo{volume}{42}, \bibinfo{pages}{1895--1976}.
\bibitem[{Kawahara et~al.(2002)Kawahara, Chung and
  Kawaji}]{kawahara2002investigation}
\bibinfo{author}{Kawahara, A.}, \bibinfo{author}{Chung, P.Y.},
  \bibinfo{author}{Kawaji, M.}, \bibinfo{year}{2002}.
\newblock \bibinfo{title}{Investigation of two-phase flow pattern, void
  fraction and pressure drop in a microchannel}.
\newblock \bibinfo{journal}{International journal of multiphase flow}
  \bibinfo{volume}{28}, \bibinfo{pages}{1411--1435}.
\bibitem[{Kawaji and Chung(2004)}]{kawaji2004adiabatic}
\bibinfo{author}{Kawaji, M.}, \bibinfo{author}{Chung, P.Y.},
  \bibinfo{year}{2004}.
\newblock \bibinfo{title}{Adiabatic gas--liquid flow in microchannels}.
\newblock \bibinfo{journal}{Microscale Thermophysical Engineering}
  \bibinfo{volume}{8}, \bibinfo{pages}{239--257}.
\bibitem[{Li et~al.(2000)Li, Firoozabadi et~al.}]{li2000}
\bibinfo{author}{Li, K.}, \bibinfo{author}{Firoozabadi, A.}, et~al.,
  \bibinfo{year}{2000}.
\newblock \bibinfo{title}{Phenomenological modeling of critical condensate
  saturation and relative permeabilities in gas/condensate systems}.
\newblock \bibinfo{journal}{Spe Journal} \bibinfo{volume}{5},
  \bibinfo{pages}{138--147}.
\bibitem[{Lohrenz et~al.(1964)Lohrenz, Bray, Clark
  et~al.}]{lohrenz1964calculating}
\bibinfo{author}{Lohrenz, J.}, \bibinfo{author}{Bray, B.G.},
  \bibinfo{author}{Clark, C.R.}, et~al., \bibinfo{year}{1964}.
\newblock \bibinfo{title}{Calculating viscosities of reservoir fluids from
  their compositions}.
\newblock \bibinfo{journal}{Journal of Petroleum Technology}
  \bibinfo{volume}{16}, \bibinfo{pages}{1--171}.
\bibitem[{Mazarei et~al.(2019)Mazarei, Davarpanah, Ebadati and
  Mirshekari}]{mazarei2019feasibility}
\bibinfo{author}{Mazarei, M.}, \bibinfo{author}{Davarpanah, A.},
  \bibinfo{author}{Ebadati, A.}, \bibinfo{author}{Mirshekari, B.},
  \bibinfo{year}{2019}.
\newblock \bibinfo{title}{The feasibility analysis of underground gas storage
  during an integration of improved condensate recovery processes}.
\newblock \bibinfo{journal}{Journal of Petroleum Exploration and Production
  Technology} \bibinfo{volume}{9}, \bibinfo{pages}{397--408}.
\bibitem[{Mohammadi et~al.(1990)Mohammadi, Sorbie, Danesh, Peden
  et~al.}]{mohammadi1990}
\bibinfo{author}{Mohammadi, S.}, \bibinfo{author}{Sorbie, K.},
  \bibinfo{author}{Danesh, A.}, \bibinfo{author}{Peden, J.}, et~al.,
  \bibinfo{year}{1990}.
\newblock \bibinfo{title}{Pore-level modelling of gas-condensate flow through
  horizontal porous media}, in: \bibinfo{booktitle}{SPE Annual Technical
  Conference and Exhibition}, \bibinfo{organization}{Society of Petroleum
  Engineers}.
\bibitem[{Momeni et~al.(2017)Momeni, Dadvar, Hekmatzadeh and
  Dabir}]{momeni20173d}
\bibinfo{author}{Momeni, A.}, \bibinfo{author}{Dadvar, M.},
  \bibinfo{author}{Hekmatzadeh, M.}, \bibinfo{author}{Dabir, B.},
  \bibinfo{year}{2017}.
\newblock \bibinfo{title}{3d pore network modeling and simulation for dynamic
  displacement of gas and condensate in wellbore region}.
\newblock \bibinfo{journal}{International Journal of Multiphase Flow}
  \bibinfo{volume}{97}, \bibinfo{pages}{147--156}.
\bibitem[{Mott et~al.(2000)Mott, Cable, Spearing et~al.}]{mott2000}
\bibinfo{author}{Mott, R.}, \bibinfo{author}{Cable, A.},
  \bibinfo{author}{Spearing, M.}, et~al., \bibinfo{year}{2000}.
\newblock \bibinfo{title}{Measurements and simulation of inertial and high
  capillary number flow phenomena in gas-condensate relative permeability}, in:
  \bibinfo{booktitle}{SPE annual technical conference and exhibition},
  \bibinfo{organization}{Society of Petroleum Engineers}.
\bibitem[{Nagarajan et~al.(2004)Nagarajan, Honarpour, Sampath and
  McMichael}]{nagarajan2004comparison}
\bibinfo{author}{Nagarajan, N.}, \bibinfo{author}{Honarpour, M.},
  \bibinfo{author}{Sampath, K.}, \bibinfo{author}{McMichael, D.},
  \bibinfo{year}{2004}.
\newblock \bibinfo{title}{Comparison of gas-condensate relative permeability
  using live fluid vs. model fluids}.
\newblock \bibinfo{journal}{SCA2004-09} .
\bibitem[{Peng and Robinson(1976)}]{Peng:1976}
\bibinfo{author}{Peng, D.Y.}, \bibinfo{author}{Robinson, D.B.},
  \bibinfo{year}{1976}.
\newblock \bibinfo{title}{A new two-constant equation of state}.
\newblock \bibinfo{journal}{Industrial \& Engineering Chemistry Fundamentals}
  \bibinfo{volume}{15}, \bibinfo{pages}{59--64}.
\newblock \DOIprefix\doi{10.1021/i160057a011}.
\bibitem[{Roussenac(2001)}]{roussenac2001}
\bibinfo{author}{Roussenac, B.}, \bibinfo{year}{2001}.
\newblock \bibinfo{title}{Gas Condensate Well Test Analysis}.
\newblock Master's thesis. Standford University. \bibinfo{address}{Stanford}.
\bibitem[{Santos and Carvalho(2019)}]{santos}
\bibinfo{author}{Santos, M.}, \bibinfo{author}{Carvalho, M.},
  \bibinfo{year}{2019}.
\newblock \bibinfo{title}{Pore network model for retrograde gas flow in porous
  media}.
\newblock \bibinfo{journal}{Journal of Petroleum Science and Engineering} ,
  \bibinfo{pages}{106635}\DOIprefix\doi{https://doi.org/10.1016/j.petrol.2019.106635}.
\bibitem[{Sochi(2013)}]{req}
\bibinfo{author}{Sochi, T.}, \bibinfo{year}{2013}.
\newblock \bibinfo{title}{Newtonian flow in converging-diverging capillaries}.
\newblock \bibinfo{journal}{International Journal of Modeling, Simulation, and
  Scientific Computing} \bibinfo{volume}{4}, \bibinfo{pages}{1350011}.
\newblock \DOIprefix\doi{10.1142/S1793962313500116}.
\bibitem[{Wang and Mohanty(1999)}]{wang1999}
\bibinfo{author}{Wang, X.}, \bibinfo{author}{Mohanty, K.},
  \bibinfo{year}{1999}.
\newblock \bibinfo{title}{Critical condensate saturation in porous media}.
\newblock \bibinfo{journal}{Journal of colloid and interface science}
  \bibinfo{volume}{214}, \bibinfo{pages}{416--426}.
\bibitem[{Wang et~al.(2000)Wang, Mohanty et~al.}]{wang2000}
\bibinfo{author}{Wang, X.}, \bibinfo{author}{Mohanty, K.K.}, et~al.,
  \bibinfo{year}{2000}.
\newblock \bibinfo{title}{Pore-network model of flow in gas/condensate
  reservoirs}.
\newblock \bibinfo{journal}{SPE Journal} \bibinfo{volume}{5},
  \bibinfo{pages}{426--434}.
\bibitem[{Weinaug and Katz(1943)}]{WK}
\bibinfo{author}{Weinaug, C.F.}, \bibinfo{author}{Katz, D.L.},
  \bibinfo{year}{1943}.
\newblock \bibinfo{title}{Surface tensions of methane-propane mixtures}.
\newblock \bibinfo{journal}{Industrial \& Engineering Chemistry}
  \bibinfo{volume}{35}, \bibinfo{pages}{239--246}.
\newblock \DOIprefix\doi{10.1021/ie50398a028}.
\bibitem[{Zhang et~al.(2020)Zhang, Fan and Zhao}]{zhang2020investigation}
\bibinfo{author}{Zhang, A.}, \bibinfo{author}{Fan, Z.}, \bibinfo{author}{Zhao,
  L.}, \bibinfo{year}{2020}.
\newblock \bibinfo{title}{An investigation on phase behaviors and displacement
  mechanisms of gas injection in gas condensate reservoir}.
\newblock \bibinfo{journal}{Fuel} \bibinfo{volume}{268},
  \bibinfo{pages}{117373}.

\end{thebibliography}







\end{document}